\documentclass[%
    reprint,
    superscriptaddress,
     amsmath,amssymb,
     aps,
]{revtex4-2}

\usepackage{amsthm}
\usepackage{mathtools}
\usepackage{physics}
\usepackage{xcolor}
\usepackage{graphicx}
\usepackage{adjustbox}
\usepackage{placeins}
\usepackage[T1]{fontenc}
\usepackage{lipsum}
\usepackage{csquotes}
\usepackage{multirow}

\usepackage{titlesec}
\titlespacing\section{-5pt}{10pt plus 4pt minus 2pt}{2pt plus 2pt minus 2pt}

\begin{document}
\title{Randomized compiling for scalable quantum computing on a noisy superconducting quantum processor}

\author{Akel Hashim}
    \thanks{These authors contributed equally to this work. Correspondence should be addressed to A.H.:         \href{mailto:ahashim@berkeley.edu}{ahashim@berkeley.edu}}
    \affiliation{Quantum Nanoelectronics Laboratory, Department of Physics, University of California at Berkeley, Berkeley, CA 94720, USA}
    \affiliation{Graduate Group in Applied Science and Technology, University of California at Berkeley, Berkeley, CA 94720, USA}
\author{Ravi K. Naik}
    \thanks{These authors contributed equally to this work. Correspondence should be addressed to A.H.:         \href{mailto:ahashim@berkeley.edu}{ahashim@berkeley.edu}}
    \affiliation{Quantum Nanoelectronics Laboratory, Department of Physics, University of California at Berkeley, Berkeley, CA 94720, USA}
\author{Alexis Morvan}
    \affiliation{Quantum Nanoelectronics Laboratory, Department of Physics, University of California at Berkeley, Berkeley, CA 94720, USA}
    \affiliation{Computational Research Division, Lawrence Berkeley National Lab, Berkeley, CA 94720, USA}
\author{Jean-Loup Ville}
    \affiliation{Quantum Nanoelectronics Laboratory, Department of Physics, University of California at Berkeley, Berkeley, CA 94720, USA}
\author{Bradley Mitchell}
    \affiliation{Quantum Nanoelectronics Laboratory, Department of Physics, University of California at Berkeley, Berkeley, CA 94720, USA}
    \affiliation{Computational Research Division, Lawrence Berkeley National Lab, Berkeley, CA 94720, USA}
\author{John Mark Kreikebaum}
    \affiliation{Quantum Nanoelectronics Laboratory, Department of Physics, University of California at Berkeley, Berkeley, CA 94720, USA}
    \affiliation{Materials Sciences Division, Lawrence Berkeley National Lab, Berkeley, CA 94720, USA}
\author{Marc Davis}
    \affiliation{Computational Research Division, Lawrence Berkeley National Lab, Berkeley, CA 94720, USA}
\author{Ethan Smith}
    \affiliation{Computational Research Division, Lawrence Berkeley National Lab, Berkeley, CA 94720, USA}
\author{Costin Iancu}
    \affiliation{Computational Research Division, Lawrence Berkeley National Lab, Berkeley, CA 94720, USA}
\author{Kevin P. O’Brien}
    \affiliation{Department of Electrical Engineering and Computer Science, Massachusetts Institute of Technology, Cambridge, MA, USA}
\author{Ian Hincks}
    \affiliation{Quantum Benchmark Inc., Kitchener, ON N2H 5G5, Canada}
\author{Joel J. Wallman}
    \affiliation{Quantum Benchmark Inc., Kitchener, ON N2H 5G5, Canada}
    \affiliation{Institute for Quantum Computing and Department of Applied Mathematics, University of Waterloo, Waterloo, Ontario N2L 3G1, Canada}
\author{Joseph Emerson}
    \affiliation{Quantum Benchmark Inc., Kitchener, ON N2H 5G5, Canada}
    \affiliation{Institute for Quantum Computing and Department of Applied Mathematics, University of Waterloo, Waterloo, Ontario N2L 3G1, Canada}
\author{Irfan Siddiqi}
    \affiliation{Quantum Nanoelectronics Laboratory, Department of Physics, University of California at Berkeley, Berkeley, CA 94720, USA}
    \affiliation{Computational Research Division, Lawrence Berkeley National Lab, Berkeley, CA 94720, USA}
    \affiliation{Materials Sciences Division, Lawrence Berkeley National Lab, Berkeley, CA 94720, USA}
    
\date{\today}

\begin{abstract}
\noindent
The successful implementation of algorithms on quantum processors relies on the accurate control of quantum bits (qubits) to perform logic gate operations. In this era of noisy intermediate-scale quantum (NISQ) computing, systematic miscalibrations, drift, and crosstalk in the control of qubits can lead to a coherent form of error which has no classical analog. Coherent errors severely limit the performance of quantum algorithms in an unpredictable manner, and mitigating their impact is necessary for realizing reliable quantum computations. Moreover, the average error rates measured by randomized benchmarking and related protocols are not sensitive to the full impact of coherent errors, and therefore do not reliably predict the global performance of quantum algorithms, leaving us unprepared to validate the accuracy of future large-scale quantum computations. Randomized compiling is a protocol designed to overcome these performance limitations by converting coherent errors into stochastic noise, dramatically reducing unpredictable errors in quantum algorithms and enabling accurate predictions of algorithmic performance from error rates measured via cycle benchmarking. In this work, we demonstrate significant performance gains under randomized compiling for the four-qubit quantum Fourier transform algorithm and for random circuits of variable depth on a superconducting quantum processor. Additionally, we accurately predict algorithm performance using experimentally-measured error rates. Our results demonstrate that randomized compiling can be utilized to leverage and predict the capabilities of modern-day noisy quantum processors, paving the way forward for scalable quantum computing.
\end{abstract}

\maketitle

\section{Introduction}
\noindent
The accuracy of quantum algorithms is limited by different types of errors. Interactions between qubits and the surrounding environment result in incoherent (i.e., non-unitary/irreversible) errors, leading to purity-decreasing processes such as the decoherence of a quantum state. In contrast, systematic imperfections in qubit control (e.g., detuning and calibration errors) and crosstalk on multi-qubit processors result in coherent (i.e., unitary/reversible) errors, which are purity-preserving and thus do not result in decoherence. For single qubits, coherent errors manifest as an unwanted unitary rotation by an angle $\epsilon$, $U(\mathbf{\hat{n}}, \epsilon) = e^{-i\epsilon\mathbf{\hat{n} \cdot \boldsymbol\sigma}/2}$, where $\mathbf{\hat{n}}$ is the axis of rotation and $\boldsymbol\sigma$ the Pauli vector. Common methods for measuring the average error rate $r(\mathcal{E})$ of quantum gates, such as randomized benchmarking \cite{emerson2005scalable, knill2008randomized, dankert2009exact, magesan2011scalable} (RB), define $r(\mathcal{E})$ as the average gate infidelity,
\begin{equation}
    r(\mathcal{E}) = 1 - \mathcal{F} = 1 - \int d\psi \bra{\psi} \mathcal{E}(\ket{\psi}\bra{\psi})\ket{\psi},
\end{equation}
where the fidelity $\mathcal{F}$ gives the average success probability that preparing an arbitrary pure state $\rho = \ketbra{\psi}{\psi}$ and then evolving the state through a noisy channel $\mathcal{E}(\ket{\psi}\bra{\psi})$ will return the system to the original state. While $r(\mathcal{E})$ captures the average gate infidelity, if coherent errors account for even a small fraction (e.g., $\sim 10\%$) of the average total error rate (e.g., $r(\mathcal{E}) \sim 10^{-4}$), the worst-case gate infidelity can scale as $\sqrt{r(\mathcal{E})}$ (e.g., $\sim 10^{-2}$) \cite{wallman2014randomized, sanders2015bounding, wallman2015estimating, kueng2016comparing}. Thus, the average-case and worst-case infidelities of a \textit{single computational gate} can differ by orders of magnitude in the presence of coherent errors, as has been explicitly demonstrated for the quantum processor used in this work using simultaneous gate set tomography \cite{rudinger2021experimental}. Therefore, the global impact of coherent errors is hard to predict for structured circuits \cite{proctor2020measuring}, due to both their quadratically-worse impact on gate infidelities relative to average error rates, and the potential for interference over the course of an algorithm.

In recent years, there has been growing theoretical interest in randomization methods to mitigate the problem of coherent errors in quantum computations \cite{knill2004fault, kern2005quantum, geller2013efficient, wallman2016noise, hastings2016turning, campbell2017shorter, campbell2019random, cai2020mitigating}. Experimentally, it has been shown that methods such as Pauli-frame randomization \cite{knill2004fault,kern2005quantum} (PFR) and Pauli twirling can reduce coherent errors in Clifford circuits \cite{ware2017experimental} and the two-qubit CPHASE gate \cite{song2019quantum}, respectively, as measured by gate set tomography \cite{merkel2013self, stark2014self, osti_1168946, blume2017demonstration}. Randomized compiling \cite{wallman2016noise} (RC) is a protocol for reducing coherent error rates in quantum algorithms \textit{in situ} that is more scalable and generalizable than PFR and simple Pauli twirling, and does not require \textit{a priori} knowledge of the specific error model. In this work, we demonstrate the experimental implementation of RC on a superconducting quantum processor (see Fig.~\ref{fig:Figure1}(a)). We show that RC effectively reduces and stabilizes the otherwise unpredictable impact of actual performance-limiting coherent errors in the quantum Fourier transform \cite{coppersmith2002approximate} (QFT) algorithm and in random circuits of variable depth sampled from a universal gate set. Furthermore, we accurately predict algorithm performance under RC from error rates measured in a scalable manner via cycle benchmarking \cite{erhard2019characterizing}, and show how RC performance gains are expected to improve as error rates in quantum processors continue to decrease, paving the way for more robust large-scale quantum computation.

\section{Randomized Compiling Protocol}

\begin{figure*}
    \includegraphics[width=2\columnwidth]{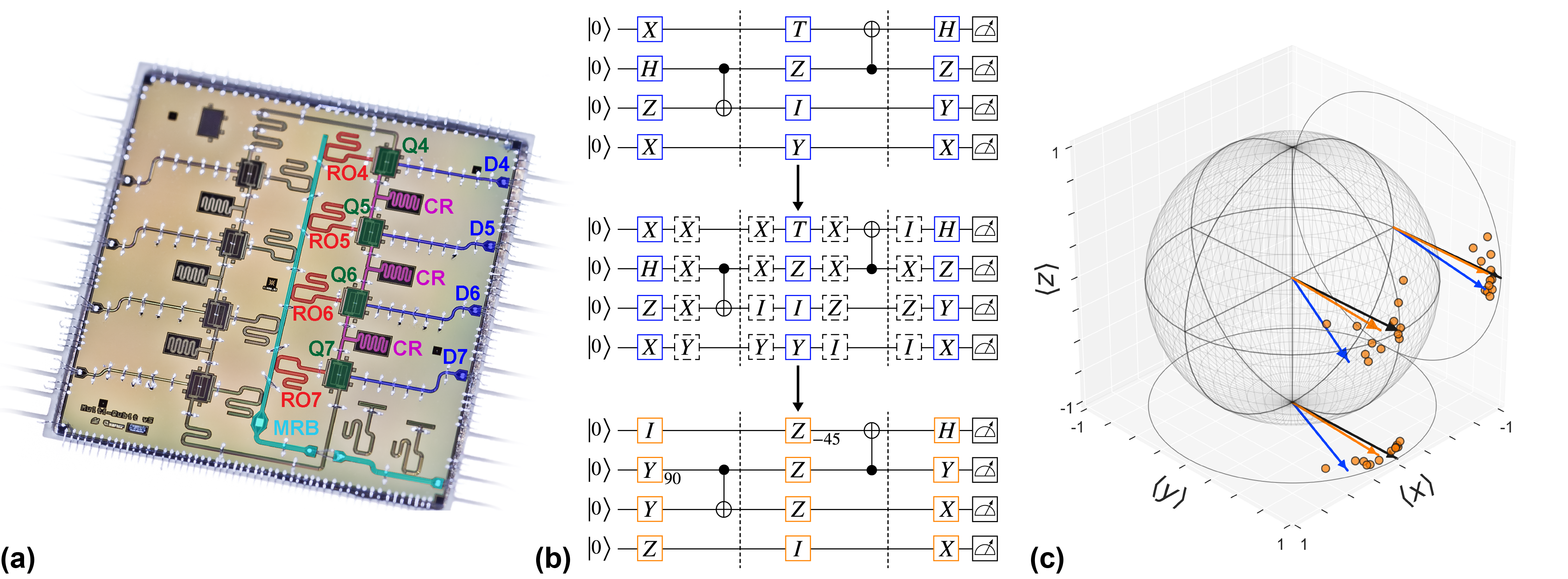}
    \caption{
        \textbf{Experimental realization of noise tailoring via randomized compiling on a superconducting quantum processor.}
        \textbf{(a)} \normalfont{False-colored micrograph of our eight-qubit superconducting quantum processor. In this work, we used four transmon qubits \cite{koch2007charge} (green) with independent microwave control lines (blue). Two-qubit cross-resonance \cite{paraoanu2006microwave, rigetti2010fully, chow2011simple, sheldon2016procedure} gates are mediated by coupling resonators (CR, purple) between nearest-neighbors. The qubits are simultaneously measured via dispersive coupling \cite{wallraff2004strong} to independent readout resonators (RO, red) coupled to a multiplexed readout bus (MRB, cyan).} 
        \textbf{(b)} \normalfont{Randomization of a quantum circuit. The bare circuit, (top) split into $K$ cycles of easy/hard gates (separated by dashed lines), (middle) is converted into a logically-equivalent circuit by inserting random single-qubit twirling gates between each easy and hard cycle, inverting them in the following cycle, (bottom) and then compiling the twirling gates into a new easy gate cycle.} 
        \textbf{(c)} \normalfont{Experimental single-qubit state-tomography results demonstrating noise tailoring: the black vector is the ideal (noiseless) final state of the qubit, but coherent errors cause an over-rotation in the measured state (blue vector). The orange vector represents the final state of the combined distribution of $N=12$ randomizations (orange data points), which has a lower purity due to the tailored stochastic noise. RC significantly mitigated the impact of coherent errors, as indicated by a reduction in the TVD from $d_\textrm{TV,bare} = 0.170(8)$ to $d_\textrm{TV,RC} = 0.029(2)$ in the $\{\ket{+X}, \ket{-X}\}$ basis, from $d_\textrm{TV,bare} = 0.069(4)$ to $d_\textrm{TV,RC} = 0.060(1)$ in the $\{\ket{+Y}, \ket{-Y}\}$ basis, and from $d_\textrm{TV,bare} = 0.073(8)$ to $d_\textrm{TV,RC} = 0.008(2)$ in the $\{\ket{0}, \ket{1}\}$ (computational) basis.} }
    \label{fig:Figure1}
\end{figure*}

\noindent
RC tailors coherent errors into a stochastic noise channel by combining the results of many logically-equivalent circuits. By inserting and compiling random single-qubit (virtual) twirling gates into a circuit in a way that preserves the overall unitary operation, RC creates a set of ``randomized'' circuits that are logically equivalent to the original ``bare'' circuit, without increasing circuit depth. Any bare circuit composed of $K$ cycles of interleaved single-qubit ``easy'' gates and two-qubit ``hard'' gates can be randomized using the following method, shown in Fig.~\ref{fig:Figure1}(b):
\begin{enumerate}
    \item Conjugate each round of easy gates $C_k$ by a twirling gate $T_k$ randomly sampled from a set $\mathcal{T}$ and an inverting operator $T^c_{k-1}$: $C_k \rightarrow T_k C_k T^c_{k-1}$, where $T^c_{k-1}$ is chosen to undo the twirling gate that was inserted in the previous cycle when commuted through the hard gate cycle $G_k$: $T^c_{k} = G_k T^\dagger_k G^\dagger_k$.
    \item Compile the original single-qubit gates and twirling gates into new easy gate cycles: $C_k' = T_k C_k T^c_{k-1}$.
\end{enumerate}
Typically, $\mathcal{T}$ is chosen to be the set of tensor products of single-qubit Paulis, with the edge terms $T^c_0$ and $T_K$ set to the identity gate, so that if the hard gates are all Clifford gates, then the correction gates will also lie in $\mathcal{T}$, and we need place no restriction on the types of allowed easy gates. Therefore, RC is efficiently compatible with universal quantum computation. In step 2, the new randomized circuit is logically equivalent to the original bare circuit and has the same number of elementary gates. Generating many ($N$) logically-equivalent randomizations of a bare circuit requires very low classical overhead and can be efficiently done before runtime. However, certain hardware platforms (e.g., superconducting circuits) may be better equipped to measure large $N$ than others (e.g., trapped ions) due to faster gate times. 

By measuring each randomization $m/N$ times and computing the union of all $N$ results, we obtain an equivalent statistical distribution for a circuit measured $m$ times in which coherent errors in each computational cycle (except the last) have been averaged into Pauli channels (e.g., random phase- and bit-flips),
\begin{equation}\label{stochastic_channel}
    \mathcal{E}(\rho) = \sum_{P \in \mathbb{P}^{\otimes n}} c_P P \rho P^\dagger,
\end{equation}
where $\rho$ is an $n$-qubit density matrix, $\mathbb{P}^{\otimes n} = \{I, X, Y, Z\}^{\otimes n}$ the set of $4^{n}$ generalized Pauli operators, and $c_P$ the relative probability of an error due to $P$. Tailoring coherent errors into stochastic Pauli noise has several major advantages: (1) the tailored noise completely suppresses off-diagonal terms in the error process resulting from coherent errors (in the limit of perfectly-implemented Pauli twirling; see Supplemental Material), reducing the overall error rate per computational gate cycle. (2) Stochastic Pauli errors have a finite probability of occurring in each gate cycle and only grow linearly with circuit depth (in the small error limit), in contrast to coherent errors which can accumulate up to quadratically with circuit depth in the rare instances in which there is complete constructive interference; thus, RC stabilizes the error rate during algorithms by breaking up the coherent accumulation of unitary errors. (3) Stochastic noise has dramatically lower worst-case error rates than coherent errors occurring at the same average rate, as defined via the diamond norm (see Supplemental Material), and can be directly estimated via randomized benchmarks in order to compare experimental error rates to fault-tolerant error rate thresholds based on Pauli noise. And (4), known fault-tolerant thresholds for stochastic noise \cite{aliferis2007subsystem, aliferis2008accuracy} are orders of magnitude higher than the threshold for generic local errors \cite{aharonov2008fault} (for example, those due to coherent errors), potentially enabling fault-tolerant error correction with error rates comparable to those already achieved in modern-day experiments.

To demonstrate noise tailoring via RC, we performed state tomography on a single qubit (Q7) after 50 random gates (see Appendix A), as shown in Fig.~\ref{fig:Figure1}(c). We find that coherent errors cause a net over-rotation in the measured state compared to the ideal (noiseless) final state. When RC is applied, each randomization results in a different net coherent error; however, the combined result is more aligned with the ideal vector. The state fidelity $\mathcal{F} = 0.862$ and purity $\gamma = 0.938$ for the bare result, and $\mathcal{F} = 0.879$ and $\gamma = 0.881$ for the RC result. While the fidelities are comparable, the rotation error in the bare result has been tailored into stochastic noise under RC, as the fidelity and purity of the RC result are approximately equal in magnitude.

To evaluate the efficacy of RC, we assess algorithmic performance by the total variation distance (TVD), a standard metric for the statistical distance between two probability distributions and a relevant measure in quantum supremacy experiments \cite{zhong2020quantum}:
\begin{equation}
    d_\textrm{TV}(\mathcal{P}, \mathcal{P}_\textrm{ideal}) = \frac{1}{2} \sum_{x \in X} |\mathcal{P}(x) - \mathcal{P}_\textrm{ideal}(x)|,
\end{equation}
where $\mathcal{P}_\textrm{ideal}(x)$ is the ideal probability of measuring a bit string $x$ in a set of possible bit strings $X$, and $\mathcal{P}(x)$ is the experimentally-measured probability. The TVD is a basis-dependent metric which determines the \textit{probability of obtaining an incorrect solution}, with 0 (1) indicating that the correct distribution of bit strings is always (never) measured. Thus, improvements in algorithmic performance equate to lower TVDs, as exemplified by the observed reduction from $d_\textrm{TV,bare} = 0.073(8)$ to $d_\textrm{TV,RC} = 0.008(2)$ for the single-qubit results in Fig.~\ref{fig:Figure1}(c), as measured in the computational basis.

More generally, in the presence of coherent errors the TVD can be as large as $d_\textrm{TV}(\mathcal{P}, \mathcal{P}_\textrm{ideal}) \leq \sqrt{r(\mathcal{E})}\sqrt{d(d+1)}$, but under RC it is instead upper-bounded directly by the average error rate $r(\mathcal{E})$,
\begin{equation}
    d_\textrm{TV}(\mathcal{P}_{\textrm{RC}}, \mathcal{P}_\textrm{ideal}) \leq r(\mathcal{E})\frac{d+1}{d},
\end{equation}
which is quadratically lower in $r(\mathcal{E})$ and does not scale with the dimension $d=2^n$ (for $n$ qubits). Thus, RC provides a general error reduction from $\sqrt{r(\mathcal{E})} \longrightarrow r(\mathcal{E})$. While many NISQ applications (e.g., variational quantum algorithms (VQAs)) depend on measured expectation values as opposed to the TVD, the TVD upper bounds the absolute error of all expectation values measured in the same basis (see Supplemental Material), therefore a small TVD under RC guarantees a small error in any expectation value estimated from the same probability distribution. Finally, the TVD is closely related to the trace distance between $\rho$ and $\rho_\textrm{ideal}$, $D(\rho, \rho_\textrm{ideal}) = \frac{1}{2} \Tr |\rho - \rho_\textrm{ideal}|$. Therefore, if $\rho$ and $\rho_\textrm{ideal}$ are close in trace distance, then any measurement of the two density matrices in the same basis will result in probability distributions that are close in total variation distance.

\section{Cycle Error Reconstruction}

\begin{figure*}
    \includegraphics[width=2\columnwidth]{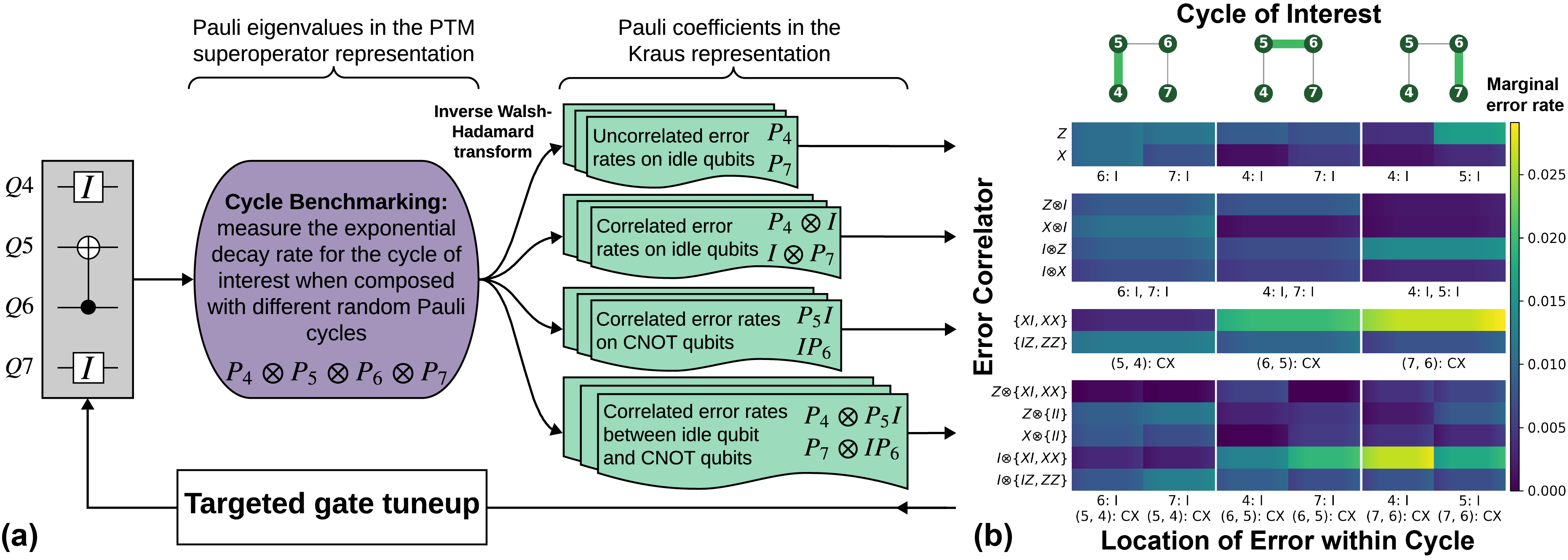}
    \caption{
        \textbf{Cycle error reconstruction of the tailored noise under cycle benchmarking.}
        \textbf{(a)} \normalfont{Schematic of the process by which single- and two-body gate errors can be reconstructed using targeted CB measurements of a parallel gate cycle (e.g., CNOT between Q5 and Q6, with Q4 and Q7 idling), providing detailed information about the Pauli errors occurring during the cycle, as shown in \textbf{b}.} 
        \textbf{(b)} \normalfont{Cycle error reconstruction results of four-qubit cycles containing a single CNOT gate and identity gates on the spectator qubits. The y-axis (x-axis) labels the type of error (where the error occurs), and the color (gradient) indicates the marginal error rate from all Pauli contributions (95\% confidence interval). The first and third rows of subplots show single-body errors, the second row of subplots shows correlated errors between idling qubits, and the last row of subplots shows correlated errors between an idling spectator qubit and a CNOT on a different pair. Curly brackets indicate error types that cannot be distinguished due to degeneracies, and any rows in which all errors are below 30\% of the maximum value have been omitted for clarity. This detailed information can be used to perform targeted gate tuneup to address the most harmful errors. The residual errors in our system are broadly distributed among many pathways, so any further targeted tuneup will come with diminishing returns.} }
    \label{fig:Figure2}
\end{figure*}

\noindent
Because RC tailors noise within a quantum circuit, it is useful to benchmark the tailored noise within our multi-qubit system. To do so, we use cycle benchmarking \cite{erhard2019characterizing} (CB), a scalable protocol that measures errors affecting all qubits during parallel gate cycles. CB has several advantages over common RB protocols: firstly, CB enables one to benchmark composite gate cycles, giving more accurate estimates of cycle performance in the context of parallel operations within a quantum circuit. Secondly, CB can capture the impact of coherent errors on idling spectator qubits, such as those not explicitly involved in an entangling gate \cite{krinner2020benchmarking}. Finally, like RC, CB tailors noise using Pauli twirling, therefore the effective noise of any cycle under CB is equal to the tailored noise under RC (Eq.~\ref{stochastic_channel}), enabling accurate predictions of algorithmic performance under RC via CB process infidelities.

In this work, we leveraged a cycle error reconstruction (CER) protocol \cite{erhard2019characterizing, flammia2019efficient} based on targeted CB measurements to produce an error map of the Pauli error rates in our four-qubit system. Fig.~\ref{fig:Figure2}(a) outlines the process by which CB can be used to reconstruct single- and two-body gate errors that occur during any hard gate cycle involving a single CNOT gate and identity gates on the spectator qubits. The error rates in CER are the coefficients $c_P$ of a Pauli channel in the Kraus representation (Eq.~\ref{stochastic_channel}), which can be reconstructed from the Pauli eigenvalues (in the Pauli transfer matrix (PTM) superoperator representation) measured by CB via linear inversion using an inverse Walsh-Hadamard transform \cite{flammia2019efficient}. Using CER, we identify the major sources of errors in our system and compensate the most harmful effects with targeted decoupling pulses or virtual phase gates  (see Supplemental Material for more details). The results plotted in Fig.~\ref{fig:Figure2}(b) show that the residual error syndromes are broadly distributed, collectively contributing to the process infidelity of each cycle and making further targeted error mitigation less fruitful. These error rates can be used to accurately predict algorithm performance under RC, as shown in Fig.~\ref{fig:Figure3}(c).

\section{Quantum Fourier Transform}
\noindent

\begin{figure*}
    \includegraphics[width=2\columnwidth]{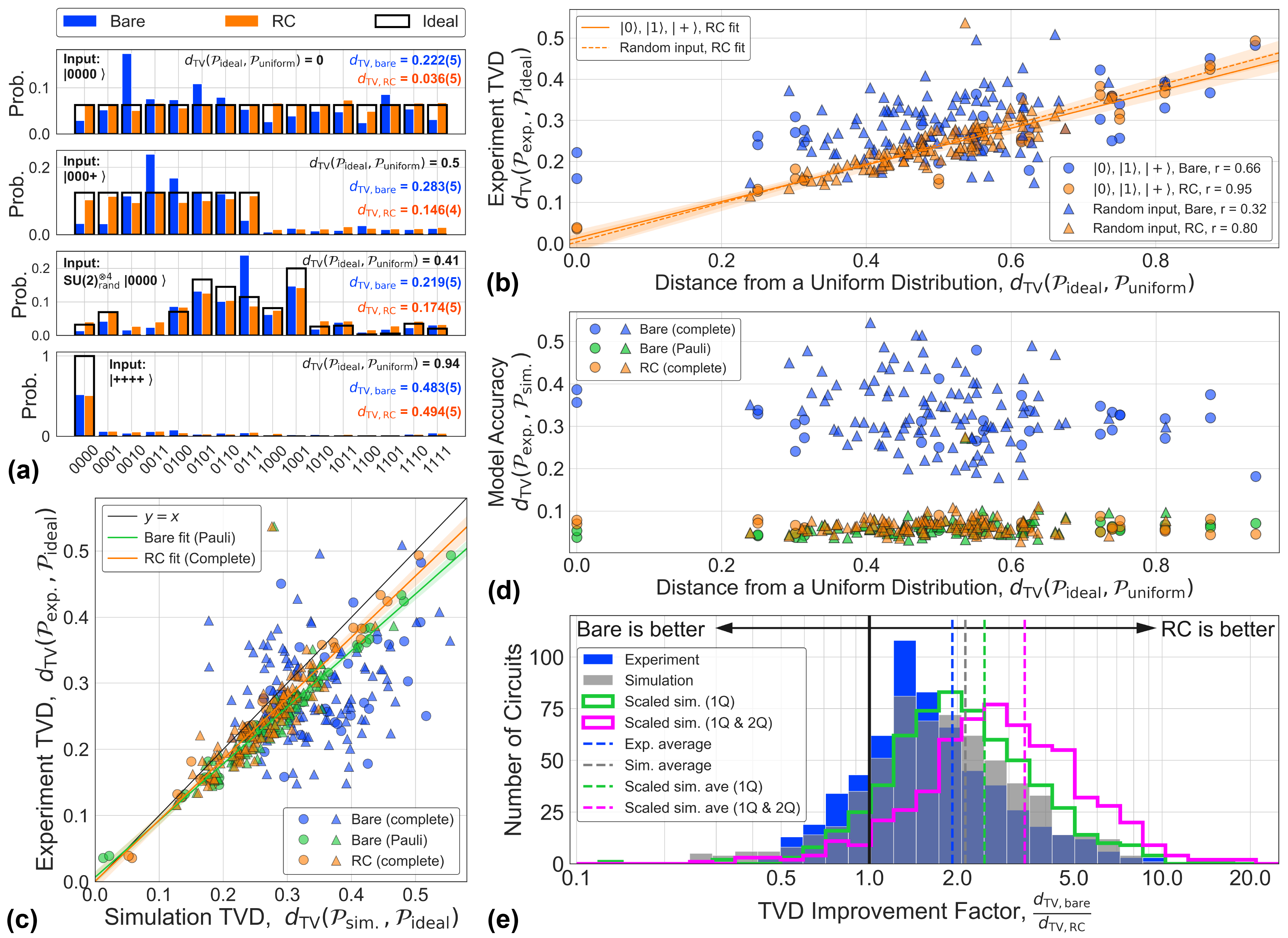}
    \caption{
        \textbf{Improving the quantum Fourier transform with randomized compiling.}
        \textbf{(a)} \normalfont{Measured probability distributions for the QFT applied to $\ket{0000}$, $\ket{000+}$, $\ket{++++}$, and a random input state ($SU(2)_\textrm{rand}^{\otimes 4}\ket{0000}$).} 
        \textbf{(b)} \normalfont{Bare and RC TVDs for all four-qubit QFT results, as a function of distribution uniformity of the ideal results. RC provides more improvement as the resultant distribution spans more basis states ($d_\textrm{TV}(\mathcal{P}_\textrm{ideal}, \mathcal{P}_\textrm{uniform}) \longrightarrow 0$). Pearson $r$ values listed in the legend quantify the correlation strength of each data set, justifying linear fits for the RC data (transparent bands indicate the 95\% confidence intervals).}
        \textbf{(c)} \normalfont{Experimental vs. simulated TVDs from two models of our system based on the Pauli error rates in Fig.~\ref{fig:Figure2}(b); the complete model includes coherent errors. The blue (orange) markers denote the bare (RC) circuits simulated with the complete model, and the green markers denote the bare circuits simulated with the Pauli model.}
        \textbf{(d)} \normalfont{Accuracy of the two models compared to experimental results. The bare circuits simulated with the Pauli model are plotted against the experimental RC results in \textbf{c}, which are also used to compute the model accuracy in \textbf{d}. For \textbf{c} and \textbf{d}, the circular (triangular) markers denote the $\{ \ket{0}$, $\ket{1}$, $\ket{+} \}$ input (random input) states.}
        \textbf{(e)} \normalfont{Summary of the improvement under RC for all two-, three-, and four-qubit random input QFT results, showing good agreement between experiment (blue) and theory (grey). Simulations in which single-qubit (green) and two-qubit (pink) error rates have been scaled down by a factor of 10 suggest that RC performance increases as error rates decrease. (Error bars on the TVD ($\mathcal{O}(10^{-3})$) for \textbf{b}, \textbf{c}, and \textbf{d} are smaller than the markers.)} }
    \label{fig:Figure3}
\end{figure*}

RC can be applied to any gate-based quantum algorithm, including those at the heart of many quantum applications, like the quantum Fourier transform (QFT). Here, we utilized a synthesis algorithm \cite{davis2019heuristics} to numerically approximate the four-qubit QFT circuit unitary in order to reduce the CNOT count to $K = 13$ for our linear connectivity (see circuit diagram in Supplemental Material). Much like the classical discrete Fourier transform, the QFT maps singular inputs (e.g., $\ket{0000}$) into uniform distributions, and maps superposition states (e.g., $\ket{++++}$) into singular distributions. To measure the performance of RC for different resultant probability distributions, we applied the QFT to various single-qubit product states involving permutations of Pauli basis states $\{ \ket{0}$, $\ket{1}$, $\ket{+} \}$, as well as random separable input states ($SU(2)_\textrm{rand}^{\otimes 4}\ket{0000}$); see Fig.~\ref{fig:Figure3}(a) for several examples of the measured distributions.

In Figs.~\ref{fig:Figure3}(a/b), we show that the relative RC performance is best (equivalent) when the algorithm generates a uniform (singular) distribution across all measurement basis states. This is due to the basis-dependence of the TVD: if the target state is an eigenstate of the measurement basis, the raw probabilities will not be sensitive to off-diagonal terms in the error process resulting from coherent errors, so RC provides no overall benefit. Therefore, distribution uniformity is a good proxy for the susceptibility of the target state to coherent errors with respect to the measurement basis, and is thus correlated with improvement under RC. We quantify the \textit{distance from a uniform distribution} by computing the TVD of each ideal probability distribution with the uniform distribution in $d = 2^{n}$ dimensions for $n$ qubits, $d_\textrm{TV}(\mathcal{P}_\textrm{ideal}, \mathcal{P}_\textrm{uniform})$, which is 0 (maximized) when $\mathcal{P}_\textrm{ideal}$ is uniform (singular). In Fig.~\ref{fig:Figure3}(b), the bare and RC TVDs are plotted as a function of $d_\textrm{TV}(\mathcal{P}_\textrm{ideal}, \mathcal{P}_\textrm{uniform})$ for all four-qubit QFT results. For singular input states ($\ket{0000}$ or $\ket{1111}$), RC significantly reduces the TVD, but for a superposition input state ($\ket{++++}$), the bare and RC TVDs are approximately equal. We compute the Pearson correlation coefficient $r$ to quantify the correlation strength between the experiment TVD and $d_\textrm{TV}(\mathcal{P}_\textrm{ideal}, \mathcal{P}_\textrm{uniform})$, where +1 (-1) indicates exact positive (negative) correlation and 0 implies no linear correlation. The RC results are strongly correlated ($r = 0.95 \; (0.80)$ for basis (random) inputs) compared to the bare results ($r = 0.66 \; (0.32)$ for basis (random) inputs), underscoring the stability and predictability of RC compared to non-randomized circuits.

In Fig.~\ref{fig:Figure3}(c), we predict the TVD performance of the QFT using two models: (1) a Pauli model of our system consisting of the Pauli error rates extracted from the CER results in Fig.~\ref{fig:Figure2}(b); (2) a complete model of our system that includes coherent errors which, under simulated CER, produces approximately equal error rates as the experimental results in Fig.~\ref{fig:Figure2}(b). The complete model is generated by finding a desired completely positive and trace-preserving (CPTP) map for a fixed unitarity \cite{wallman2015estimating} (i.e., fixed fraction of the total error rate due to coherent errors) which has been minimized with respect to the experimental CER results (see Supplemental Material for more details). Using the Pauli model, we simulate the bare QFT circuits and compare these results to the experimental RC TVDs. Using the complete model, we simulate both the bare and RC circuits and compare the results to their respective experimental TVDs. We see excellent agreement between experiment and simulation for the bare circuits simulated with the Pauli model and the RC circuits simulated with the complete model, but unreliable predictability for the bare circuits simulated with the complete model. This tells us that we can accurately predict the results of an algorithm \textit{a priori} with CER error rates using (1) a Pauli model, as long as the experimental circuit is performed using RC, and (2) using a complete model, as long as both the simulated and experimental circuits are performed using RC. In Fig.~\ref{fig:Figure3}(d), we validate the model accuracy by computing the TVD of the experimental results with the simulated results. Almost all of the simulated bare (Pauli) and RC (complete) results are accurate to within 10\% with respect to the experimental RC results, but the accuracy of the simulated bare (complete) results are much worse due to the difficulty in modeling the complex interplay of coherent errors. By utilizing RC in conjunction with CB/CER, we close the gap between the circuit performance predicted from benchmarking diagnostics and experimental results.

In practice, the input states to quantum algorithms will not be known \textit{a priori}, such as when the QFT is used in Shor's algorithm \cite{shor1999polynomial}. While unknown inputs are not guaranteed to be random, we use random inputs as a proxy for when the QFT is used as a subroutine in other algorithms. Fig.~\ref{fig:Figure3}(b) shows that when the QFT is applied to random input states, most of the results are improved under RC. A histogram of the TVD improvement for two-, three-, and four-qubit random input QFT results can be seen in Fig.~\ref{fig:Figure3}(e) (two- and three-qubit QFT results are provided in the Supplemental Material), showing that the vast majority of circuits ($> \text{81\%}$) are improved under RC by an average of $d_\textrm{TV,bare} / d_\textrm{TV,RC} \approx 1.9$. Here, we include two- and three-qubit results in order to summarize the RC QFT performance using a larger sample size drawn from systems that include differing error rates. In the rare instances in which coherent errors in a circuit benignly cancel, RC can hurt performance ($d_\textrm{TV,bare} / d_\textrm{TV,RC} < 1$); however, in general, this becomes vanishingly unlikely for longer depth circuits.

While the complete model may not accurately predict the individual result of any given bare circuit, it does predict the average performance, which is possible as long as the magnitude of the unitarity in the model is correct. This can be seen by the approximate symmetry of the bare (complete) points around the $y=x$ line in Fig.~\ref{fig:Figure3}(c), and in Fig.~\ref{fig:Figure3}(e), in which the distribution of improvement under RC predicted by simulation agrees well with experiment, with an overlapping index of 0.94 (out of a maximum of 1), which quantifies the percentage that one normal distribution overlaps with another. The good agreement between experiment and theory in Fig.~\ref{fig:Figure3}(e) suggests that we can predict the average improvement under RC as error rates decrease. Included in Fig.~\ref{fig:Figure3}(e) are simulated results in which single-qubit error rates are reduced by a factor of 10, resulting in a modest improvement, and when both single- and two-qubit error rates are reduced by a factor of 10, in which case RC improves $> \text{94\%}$ of the simulated circuits by an average of $d_\textrm{TV,bare} / d_\textrm{TV,RC} \approx 3.4$. In agreement with the predictions made in Ref.~\cite{wallman2016noise}, these results demonstrate that RC is expected to provide a larger relative improvement as gate infidelities decrease (for a fixed fraction of the total error rate due to coherent errors). Therefore, as quantum processors improve and error rates decrease, we can expect RC to outperform non-randomized circuits as long as coherent errors persist.

\begin{figure*}
    \includegraphics[width=2\columnwidth]{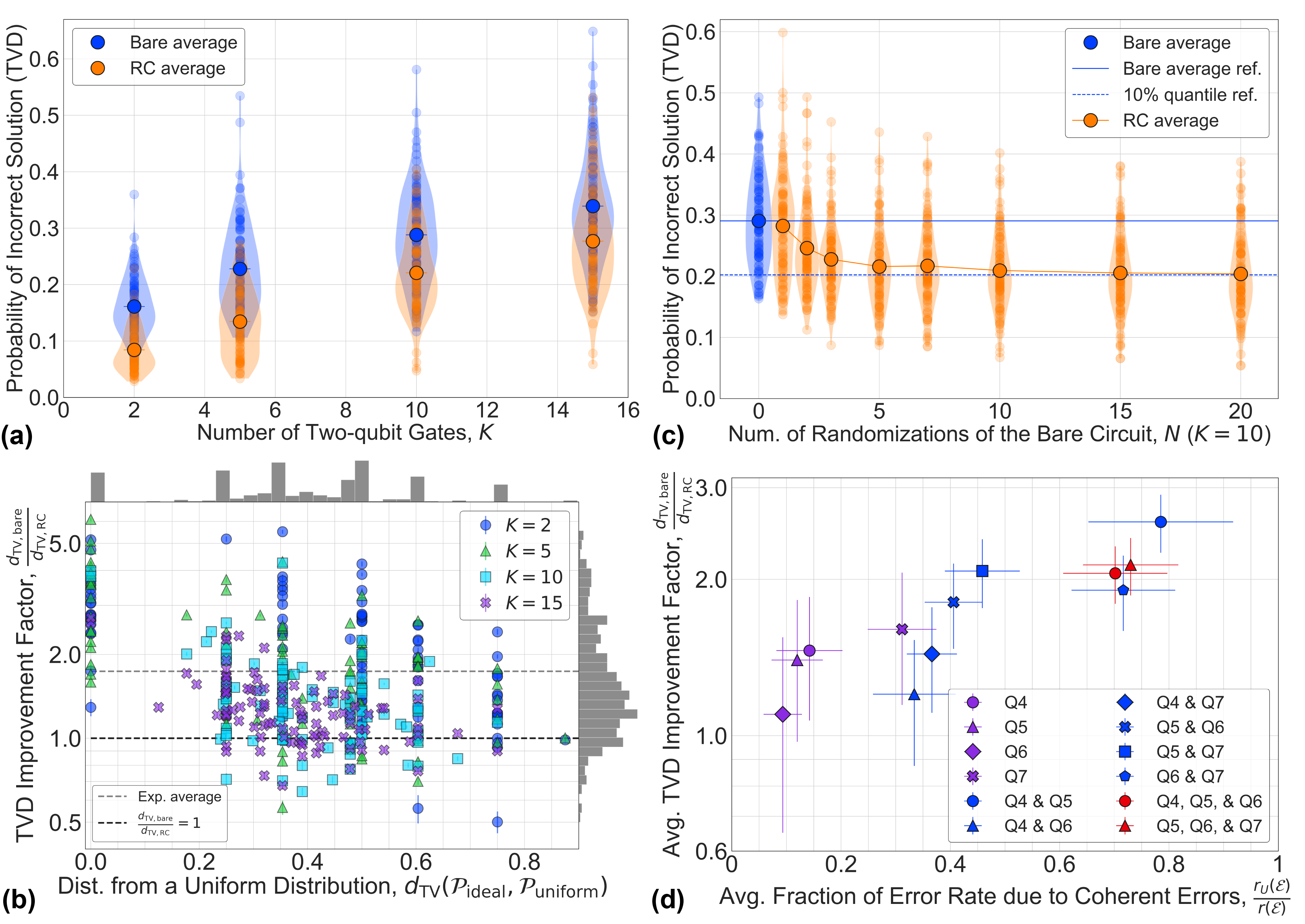}
    \caption{
        \textbf{Randomized compiling extends the computational reach with respect to circuit depth.}
        \textbf{(a)} \normalfont{Bare and RC TVDs as a function of circuit depth $K$. RC reduces the TVD on average for all circuit depths tested, allowing one to perform longer gate sequences under a fixed TVD error budget. The semi-transparent blue (orange) points indicate the TVDs of the individual random circuits (the unioned data over all $N=20$ randomizations of the corresponding bare circuits). Violin plots depict the distribution of results. The TVD error grows approximately linearly with $K$ for both bare and RC results, suggesting that the dominant benefit of RC is a reduction in the error rate per gate cycle.} 
        \textbf{(b)} \normalfont{RC TVD improvement factor for the random circuits in \textbf{a} as a function of distribution uniformity. The average improvement is $d_\textrm{TV,bare} / d_\textrm{TV,RC} \approx 1.7$.}
        \textbf{(c)} \normalfont{TVD as a function of number of randomizations, with $K=10$ fixed. The average TVD under RC converges to a value close to the 10\% quantile level (dashed line) of the non-randomized circuits for $N=20$. However, only $N=10$ randomizations are needed to converge to within 2.7\% of the $N=20$ level.}
        \textbf{(d)} \normalfont{For a fixed total error rate, RC provides a larger TVD improvement for systems with a higher fraction of coherent errors. The colored subsets listed in the legend highlight random single-qubit circuits that were performed in isolation (purple) or in parallel (blue and red). The average fraction of the total error rate due to coherent errors was quantified using measurements of RB and unitary RB under isolated or simultaneous operation.}}
    \label{fig:Figure4}
\end{figure*}

\section{Random Circuits of Variable Depth}

\noindent
To illustrate the broad applicability and generic benefits of RC for universal circuits, we demonstrate achievable performance gains for RC applied to four-qubit circuits of variable depth composed of $K$ interleaved cycles of easy/hard gates randomly sampled from a universal (Clifford + $T$) gate set (see Appendix C). As shown in Fig.~\ref{fig:Figure4}(a), RC reduces the average TVD at all circuit depths tested (with $N = 20$ randomizations for each bare circuit), demonstrating how longer-depth quantum circuits can be performed under RC given a fixed error budget in the TVD. Fig.~\ref{fig:Figure4}(a) also shows that the average TVD error grows approximately linearly with circuit depth for both the bare and RC circuits, suggesting that the dominant reason for an improvement under RC is an overall reduction in the error contribution from each gate cycle $K$, rather than a suppression of the adversarial accumulation of coherent errors (although RC can additionally provide this benefit). The relative improvement under RC is reduced at longer circuit depths, since both the bare and RC results will converge to a uniform distribution (i.e., statistical mixture) due to decoherence in the limit of large $K$. The TVD improvement under RC for all of the random circuits in Fig.~\ref{fig:Figure4}(a) is plotted in Fig.~\ref{fig:Figure4}(b) as a function of $d_\textrm{TV}(\mathcal{P}_\textrm{ideal}, \mathcal{P}_\textrm{uniform})$, with an average improvement of $d_\textrm{TV,bare} / d_\textrm{TV,RC} \approx 1.7$. Interestingly, randomly sampled circuits are not uniformly spread across $d_\textrm{TV}(\mathcal{P}_\textrm{ideal}, \mathcal{P}_\textrm{uniform})$ for small $K$; rather, they are highly concentrated at several uniformities, and only begin spreading out for larger $K$. However, RC still provides an average improvement at large $K$. Given the concentration of results in the range between $d_\textrm{TV}(\mathcal{P}_\textrm{ideal}, \mathcal{P}_\textrm{uniform}) \in [0.2, 0.8]$ in both Fig.~\ref{fig:Figure3}(b) and Fig.~\ref{fig:Figure4}(b), we would expect typical algorithms to fall within this range.

Additionally, we show that a small number of randomizations is sufficient to saturate the lowest-possible TVD under RC for a fixed circuit depth ($K=10$), plotted in Fig.~\ref{fig:Figure4}(c). After $N=20$ randomizations, the average TVD under RC converges to a value that is better than approximately 90\% of the non-randomized circuits. However, after only $N=10$ randomizations, the average RC TVD is already within 2.7\% of the $N=20$ level, highlighting the resource-efficiency of this protocol. Furthermore, we observe that a single randomization offers a slight improvement over the bare circuits. We conjecture that this is due to RC providing some level of dynamical decoupling from the adversarial accumulation of coherent errors and non-Markovian noise in the single-randomization limit, adding to the robustness to such errors already provided by randomly sampled circuits.

Finally, in Fig.~\ref{fig:Figure4}(d) we plot the average TVD improvement factor for random single-qubit circuits (at a fixed circuit depth $K=5$; see Supplemental Material) performed in isolation or in parallel as a function of the average fraction of the total error rate due to coherent errors. From these data, we see that RC provides a larger relative TVD improvement as the fraction of the total error rate due to coherent errors increases (for a fixed total error rate). We measured the average total error rate $r(\mathcal{E})$ using RB and the average error rate due to coherent errors $r_U(\mathcal{E})$ using unitary RB \cite{wallman2015estimating, dirkse2019efficient} (see Supplemental Material). Even though the average total error rate of any two results are not exactly equal, we group the data by the number of qubits performed in parallel (differentiated by color), since $r_U(\mathcal{E}) / r(\mathcal{E})$ can be more directly compared across these subsets independently. While RC performance decreases as $r_U(\mathcal{E}) / r(\mathcal{E}) \longrightarrow 0$, we note that even for single-qubit systems which are close to coherence-limited ($r_U(\mathcal{E}) / r(\mathcal{E}) \lesssim 0.1$), RC still provides an average improvement. Therefore, while the trade-off between decreased RC improvement as $r_U(\mathcal{E}) / r(\mathcal{E}) \longrightarrow 0$ (for $r(\mathcal{E})$ fixed) and increased RC improvement as $r(\mathcal{E}) \longrightarrow 0$ (for $r_U(\mathcal{E}) / r(\mathcal{E})$ fixed) will depend on each system individually, our results suggests that any system with coherent errors can benefit from RC, even those which are nearly coherence-limited.

\section{Outlook}
\noindent
In this work, we have demonstrated the promising capabilities of randomized compiling, a universal protocol in gate-based quantum computing for suppressing coherent errors that is agnostic to specific error models and hardware platforms. RC provides a strategy for mitigating complex and intractable crosstalk dynamics, extending the computational reach of noisy quantum processors. Additionally, novel error reconstruction methods using CB are well-suited to characterize the new and emergent forms of crosstalk errors seen on multi-qubit processors, and offer a method for accurately predicting error rates under RC. This improved predictability is essential for scalable quantum computing, and is necessary for comparing experimental error rates to fault tolerant thresholds. 

We believe that our methods and results have broad relevance across many experimental and theoretical efforts exploring gate-based quantum computing applications, including NISQ algorithms such as VQAs, which depend on the accurate measurement of expectation values. Additionally, while VQAs aimed at finding ground state energies of quantum systems can converge even in the presence of coherent errors, the true parameterization of the ground state wavefunction may be incorrect; we suspect that RC can facilitate faster convergence to the ground state energy in the presence of coherent errors, while also finding the best estimate of the true ground state wavefunction of the system. Finally, RC may continue to be useful in the fault-tolerant era since it is expected that, under certain conditions, coherent errors will continue to persist and remain a problem even with quantum error correction \cite{chamberland2017hard, greenbaum2017modeling}; utilizing RC will ensure that fault-tolerant error thresholds are set by stochastic errors, not coherent errors. To this end, we expect that RC is not just a stopgap measure in the NISQ era, but will continue to be a powerful technique beyond NISQ.

\section*{Acknowledgements}
\noindent
This work was supported by the Office of Advanced Scientific Computing Research, Office of Science of the U.S. Department of Energy under Contract No. DE-AC02-05CH11231. This material is based upon work supported by the U.S. Army Small Business Technology Transfer Program Office and the Army Research Office under Contract No. W911NF-19-P-0007. A.H. acknowledges financial support from the National Defense Science \& Engineering Graduate (NDSEG) Fellowship. 

\section*{Author contributions}
\noindent
J.J.W. and J.E. conceptualized the work. A.H. and R.K.N. conducted the experiments and analyzed the data. A.M., J.L.V., I.H., J.J.W., and J.E. provided useful ideas and insight into experimental work. B.M. wrote much of the software used to perform the experiments. K.P.O. designed and J.M.K. fabricated the device. M.D, E.S., and C.I. developed the software to generate the optimized circuits \cite{davis2019heuristics}. I.H. provided True-Q \cite{beale_stefanie_j_2020_3945250} software support, helped analyze the cycle error reconstruction data, and generated the model simulator. I.S. supervised all work. A.H., R.K.N., J.E., and I.S. wrote the manuscript, with input from all coauthors. 

\section*{Competing interests}
\noindent
I.H, J.J.W., and J.E. have a financial interest in Quantum Benchmark Inc. and the use of True-Q software. The other authors declare no competing interests.

\section*{Additional Information}
\noindent
Supplemental Material is available for this paper. All data are available in the manuscript, supplementary materials, or from the corresponding author upon reasonable request.

\begin{appendix}
\section*{Appendix A. Single-qubit State Tomography}
\noindent
The random circuit used for the single-qubit state tomography results in Fig.~\ref{fig:Figure1}(c) was generated by randomly sampling $K=25$ interleaved cycles of ``easy'' and ``hard'' single-qubit gates, as defined by the following gate sets: the single-qubit Clifford set, $C_{easy} = \{ \mathbf{C_1} \}$, and common non-Clifford gates, $G_{hard} = \{ X45, \; Y45, \; T=Z45 \}$. State tomography results are reconstructed by performing ensemble measurements of the same final state in the $X$, $Y$, and $Z$ bases. For the experimental results presented in Fig.~\ref{fig:Figure1}(c), 6,000 shots were taken in each measurement basis for the bare circuit. Since 12 randomizations of the bare circuit were utilized in the RC result, in order to normalize shot statistics between the bare and RC results, 500 shots were taken for each randomization in each basis.

\section*{Appendix B. Quantum Fourier Transform}
\noindent
Each bare QFT circuit was measured 10,000 times. $N=50$ randomizations were generated for each bare circuit, and each randomization was measured 200 times. All ``random'' input states were generated by applying random $SU(2)$ unitaries to each qubit independently before applying the QFT algorithm. 100 random inputs were generated for the data presented in Fig.~\ref{fig:Figure3}(b).

\section*{Appendix C. Random Circuits of Variable Depth}
\noindent
Random bare circuits were generated by randomly sampling $K$ interleaved cycles of easy and hard gates from the following gate sets: for the four-qubit circuits in Figs.~\ref{fig:Figure4}(a/b/c), $C_{easy} = \{ \mathbf{C_1}, \; X45, \allowbreak \; Y45, \; T \}$ and $G_{hard} = \{ CX=CNOT, \; CY, \; CZ \}$, where $\mathbf{C_1}$ is the single-qubit Clifford set. For the single-qubit circuits in Fig.~\ref{fig:Figure4}(d), $C_{easy} = \{ \mathbf{C_1} \}$ and $G_{hard} = \{ X45, \; Y45, \; T \}$. For each circuit depth $K$, which we define in terms of the number of two-qubit (non-Clifford) gates for the mutli-qubit (single-qubit) circuits, 100 random bare circuits were generated, and each was measured 4,000 times. $N=20$ randomizations were generated for each random bare circuit, and each randomization was measured 200 times. All $N=20$ randomizations of the corresponding bare circuits were combined to obtain an equivalent statistical distribution for a circuit measured 4,000 times. The circuit depth was fixed at $K=5$ for the single-qubit results in Fig.~\ref{fig:Figure4}(d).

For the results presented in Fig.~\ref{fig:Figure4}(c), a total of $\mathcal{N} = 20$ randomizations were generated for each bare circuit and each was measured 4,000 times. For each $N$ along the x-axis, the union over only $N \in \mathcal{N}$ randomizations were computed for each bare circuit, and then $m=4000$ shots were randomly chosen from a total of $m_\text{total} = \sum_i^N m_i$ shots (with the exception of $N = 1$). This is equivalent to throwing out $m_\text{total} - 4000$ shots at random from the unioned RC result in order to sample a smaller distribution from the full distribution.

\end{appendix}

\bibliographystyle{apsrev4-1}
\bibliography{bibliography}

\clearpage
\onecolumngrid
\setcounter{figure}{0}

\makeatletter 
\renewcommand{\thefigure}{S\@arabic\c@figure}
\renewcommand{\thetable}{S\arabic{table}}
\makeatother

\begin{center}
    \noindent{
    \bfseries \Large Supplemental Material for: \\ 
    \bigskip
    \large Randomized compiling for scalable quantum computing on a noisy superconducting quantum processor \\
    }\setlength{\parskip}{11pt}
\end{center}

\begin{center}
    \noindent
    Akel Hashim$^{1,2,\ast}$, Ravi K. Naik$^{1}$, Alexis Morvan$^{1,3}$, Jean-Loup Ville$^{1}$, Bradley Mitchell$^{1,3}$, John Mark Kreikebaum$^{1,4}$, Marc Davis$^{3}$, Ethan Smith$^{3}$, Costin Iancu$^{3}$, Kevin P. O’Brien$^{5}$, Ian Hincks$^{7}$, Joel J. Wallman$^{6,7}$, Joseph Emerson$^{6,7}$, and Irfan Siddiqi$^{1,3,4}$ \\
    \bigskip
    \textit{$^{1}$Quantum Nanoelectronics Laboratory, Dept. of Physics, University of California at Berkeley,\\Berkeley, CA 94720, USA}\\
    \textit{$^{2}$Graduate Group in Applied Science and Technology, University of California at Berkeley,\\Berkeley, CA 94720, USA}\\
    \textit{$^{3}$Computational Research Division, Lawrence Berkeley National Lab, Berkeley, CA 94720, USA}\\
    \textit{$^{4}$Materials Sciences Division, Lawrence Berkeley National Lab, Berkeley, CA 94720, USA}\\
    \textit{$^{5}$Dept. of Electrical Engineering and Computer Science, Massachusetts Institute of Technology,\\Cambridge, MA, USA}\\
    \textit{$^{6}$Quantum Benchmark Inc., Kitchener, ON N2H 5G5, Canada}\\
    \textit{$^{7}$Inst. for Quantum Computing and Dept. of Applied Mathematics, University of Waterloo,\\Waterloo, Ontario N2L 3G1, Canada}\\
    \textit{$^\ast$E-mail correspondence: ahashim@berkeley.edu}
\end{center}

\section{Experimental Setup}
\begin{figure}
\centerline{\includegraphics[width=\textwidth]{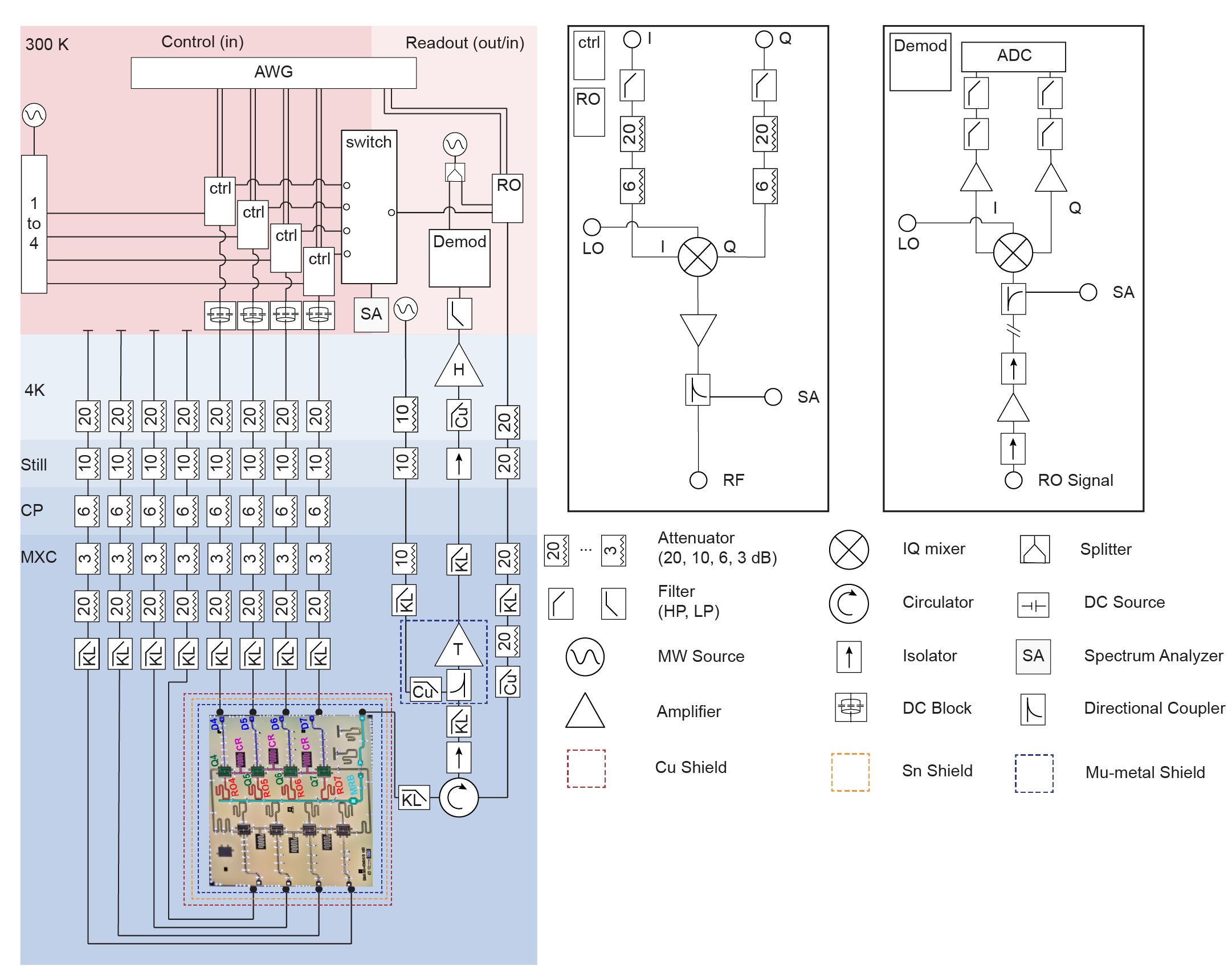}}
\caption{
    \normalfont{Experimental setup.}}
    \label{fig:Figure_S_ExperimentalSetup}
\end{figure}

\noindent
The experiments in this work were performed on a superconducting quantum processor (\texttt{AQT@LBNL Trailblazer8-v5.c2}, wafer \texttt{k180607}) cooled to $\sim$ 10 mK in a BlueFors XLD dilution refridgerator. Room-temperature and cryogenic electronics for qubit control and measurement are shown in Fig.~\ref{fig:Figure_S_ExperimentalSetup}. Qubit control pulses are generated by upconversion of IF pulses generated by a Keysight PXI Arbitrary Waveform Generator (AWG) via IQ modulation of a CW local oscillator (LO) tone, sourced by a Keysight MXG N5183B at 5.415 GHz. Both I and Q components of the IF pulses are sourced by the AWG at 1 GS/s between 66 and 261 MHz. The phase and DC offsets between the I and Q waveforms are tuned to eliminate the opposite sideband and LO leakage due to mixer nonidealities, while band-pass filtering at room-temperature reduces noise from the AWG. Measurement pulses are generated with the same AWG and are similarly upconverted with a 6.83 GHz LO tone from a separate Keysight MXG N5183B. Control pulses are passed through a DC block, and all pulses travelling to the sample are attenuated at each stage, with a further K\&L low-pass filter at the base stage.

Reflected measurement signals are redirected by a circulator to a measurement chain outfitted with superconducting coaxial cable, where they are amplified by a traveling wave parametric amplifier (TWPA) at 10 mK, a HEMT at 4K, and finally a room temperature amplifier, before being downconverted to IQ IF components with the same 6.83 GHz LO tone used for RO upconversion. The downconverted signals are then amplified and filtered to reduce high-frequency amplifier noise before being digitized at 1 GS/s by an Alazar ADC and demodulated in software. The TWPA is pumped with a CW tone, sourced by a Hittite HMC M2100 at 7.42 GHz and 16.5 dBm before being attenuated and filtered by low-pass K\&L and copper powder filters. 

\section{Device Fabrication}
\noindent
Fabrication details can be found in the Supplemental Material for Ref. \cite{blok2021quantum} and in Ref. \cite{kreikebaum2020improving}.

\section{Qubit Characterization}\label{sec_qubit_characterization}
\noindent
Parameters for the four transmon qubits \cite{koch2007charge} used in this work are given in Tables \ref{tab:sqp} and \ref{tab:tqp}. Qubit frequencies, readout frequencies, and anharmonicities are measured using standard spectroscopic methods on a vector network analyzer and/or Ramsey spectroscopy. Relaxation and coherence times are extracted by fitting decay curves to the excited state lifetime and (echoed) Ramsey measurements, respectively. Dispersive couplings $f_{ZZ}$ between neighboring qubits are measured with Ramsey experiments on one qubit conditioned on the state of the other qubit. The resonant coupling $g$ (see Table \ref{tab:tqp}) for each qubit pair is calculated from the dispersive coupling $f_{ZZ}$, the frequency detuning $\Delta$ between the qubits, and their anharmonicities $\alpha$:
\begin{equation}\label{resonant_coupling}
    2g^2 = f_{ZZ} \bigg( \frac{1}{\Delta + \abs{\alpha_T}} - \frac{1}{\Delta - \abs{\alpha_C}} \bigg)^{-1},
\end{equation}
where $C$ and $T$ refer to the control and target qubits, respectively.

Multiplexed qubit readout is performed through a common readout bus connected to independent readout resonators dispersively-coupled \cite{wallraff2004strong} to each qubit. We discriminate between the ground and excited states by fitting a Gaussian Mixture Model (GMM) to the two measurement statistics in the IQ plane, and use these fits to draw a classification boundary between the two states. Readout fidelities are subsequently determined separately for each qubit by performing ensemble measurements of the qubit prepared in the ground and excited states and classifying them using the aforementioned classification boundary. All circuit and algorithm measurements presented in this work are heralded to remove erroneous bit string results due to residual thermal excitations in the qubits. The percentage of measurements discarded due to heralding typically ranges from 5\% - 10\%, depending on the number of qubits, since each qubit is heralded independently.

\begin{table}[h]
\centering
\resizebox{0.83\textwidth}{!}{
\begin{tabular}{|l||r|r|r|r|}
\hline
{} & Q4 & Q5 & Q6 & Q7 \\
\hline
\hline
Qubit frequency [GHz] & 5.230708(5) & 5.297662(5) & 5.459108(5) & 5.633493(5) \\
Readout frequency [GHz] & 6.56381 & 6.62310 & 6.67929 & 6.73747 \\
Anharmonicity [MHz] & -273.66(1) & -273.32(1) & -270.70(1) & -267.45(1) \\
Relaxation time $T_1$ [$\mu$s] & 66(5) & 58(9) & 65(5) & 59(5) \\
Ramsey coherence time [$\mu$s] & 38(3) & 24(2) & 39(12) & 47(9) \\
Echo coherence time [$\mu$s] & 71(5) & 77(8) & 86(6) & 61(3) \\
Readout fidelity, $P(0|0)$ & 0.9969(6) & 0.9970(5) & 0.9973(6) & 0.9958(7) \\
Readout fidelity, $P(1|1)$ & 0.9872(12) & 0.9862(12) & 0.9786(15) & 0.9841(13) \\
RB infidelity, isolated & 1.2(0.06)$\times$10$^{\text{-3}}$ & 1.1(0.05)$\times$10$^{\text{-3}}$ & 1.4(0.04)$\times$10$^{\text{-3}}$ & 1.9(1)$\times$10$^{\text{-3}}$ \\
\;\; due to coherent errors & 1.6(7)$\times$10$^{\text{-4}}$ & 1.4(5)$\times$10$^{\text{-4}}$ & 1.3(5)$\times$10$^{\text{-4}}$ & 5.9(1.1)$\times$10$^{\text{-4}}$ \\
RB infidelity, simultaneous & 3.9(5)$\times$10$^{\text{-3}}$ & 6.0(8)$\times$10$^{\text{-3}}$ & 7.2(1.0)$\times$10$^{\text{-3}}$ & 5.2(5)$\times$10$^{\text{-3}}$ \\
\;\; due to coherent errors & 2.7(5)$\times$10$^{\text{-3}}$ & 4.5(8)$\times$10$^{\text{-3}}$ & 5.4(1.0)$\times$10$^{\text{-3}}$ & 3.1(5)$\times$10$^{\text{-3}}$ \\
\hline
\end{tabular}}
\caption{
    \textbf{Single-qubit parameters for the four qubits used in this work (\texttt{AQT@LBNL Trailblazer8-v5.c2}, wafer \texttt{k180607})}. The transmon frequencies were measured with Ramsey experiments, and the coherence time means and variances were extracted from repeated experiments taken over the course of a day. Ground and excited state readout fidelities are determined by repeated measurements of the qubits prepared using identity and $X$ gates, respectively, and are classified according to a GMM classification boundary in the IQ plane. Randomized benchmarking is used to measure the error rates (defined via the process infidelity) of single qubit gates performed in isolation and in parallel. Unitary RB is used to measure the error rate due to coherent errors. All uncertainties are standard deviations.}
\label{tab:sqp}
\end{table}
\begin{table}[h]
\centering
\resizebox{0.7\textwidth}{!}{
\begin{tabular}{|l||r|r|r|}
\hline
Control qubit & Q5 & Q6 & Q7 \\
\hline
Target qubit & Q4 & Q5 & Q6 \\
\hline
\hline
Dispersive coupling [kHz], $f_{ZZ}$ & 96(1) & 170.6(9) & 252.3(8) \\
Inferred resonant coupling [MHz], $g$ & 2.5 & 2.73 & 3.11 \\
CNOT Gate Duration [ns] & 135 & 147 & 174 \\
RB infidelity & 3.3(3)$\times$10$^{\text{-2}}$ & 4.2(4)$\times$10$^{\text{-2}}$ & 4.4(7)$\times$10$^{\text{-2}}$ \\
\;\; due to coherent errors & 2.0(3)$\times$10$^{\text{-2}}$ & 2.8(4)$\times$10$^{\text{-2}}$ & 2.9(7)$\times$10$^{\text{-2}}$ \\
\hline
\end{tabular}}
\caption{
    \textbf{Two-qubit parameters for \texttt{AQT@LBNL Trailblazer8-v5.c2}, wafer \texttt{k180607}.} The dispersive coupling rates are measured with Ramsey experiments on one qubit, conditioned on the state of the other qubit. The resonant coupling rates are inferred from the dispersive coupling, the frequency detuning, and the respective anharmonicities (Eq.~\ref{sec_qubit_characterization}.\ref{resonant_coupling}). CNOT gates are implemented using the cross-resonance interaction \cite{paraoanu2006microwave, rigetti2010fully, chow2011simple, sheldon2016procedure} and consist of square pulses with 30 ns cosine ramps; the total duration of the gates are listed above. Randomized benchmarking is used to measure the error rates (defined via the process infidelity) of the two-qubit gates. Unitary RB is used to measure the error rate due to coherent errors.}
\label{tab:tqp}
\end{table}

\section{Crosstalk Compensation}
\noindent
We find microwave control line crosstalk between qubits when performing single- and two-qubit gates. The dominant crosstalk is between nearest-neighbor qubits, but non-trivial crosstalk between more distant qubit exists as well. The effects of this type of crosstalk are varied, depending on the couplings between qubits and their relative transitions frequencies. Some of the errors that result from crosstalk are products of local operations, while others involve multi-qubit, entangling operations.

A simple model of crosstalk between two nearest-neighbor, coupled qubits, Q1 and Q2, can be understood as follows: consider the case where we are driving Q1 at Q1's frequency, $f_1$. In the presence of crosstalk, a drive on either line will produce a field at both qubits. Driving down Q1's line at $f_1$ produces the following Hamiltonian term:
\begin{equation}
    H_1 = (a_1 X + b_1 Y) \otimes I,
\end{equation}
where $a_1 = \cos{\phi}$, $b_1 = \sin{\phi}$, and $\phi$ the phase of the Rabi drive. However, this will also produce an effective driving term down Q2's line at $f_1$ due to crosstalk:
\begin{equation}
    H_2 = (c_1 X + d_1 Y) \otimes (a_2 Z + b_2 I),
\end{equation}
where $c_1 = \cos{\phi}$ and $d_1 = \sin{\phi}$. We assume that $f_1$ is sufficiently far from $f_2$ such that a pulse down Q2's line at $f_1$ results in an AC Stark shift on Q2 due to off-resonant driving, and thus a Hamiltonian term of the form $(Z + I)$. The full Hamiltonian is
\begin{equation}
    H = H_1 + H_2 = \alpha [(a_1 X + b_1 Y) \otimes I] + \beta [(c_1 X + d_1 Y) \otimes (a_2 Z + b_2 I)],
\end{equation}
where the ratio $\beta / \alpha$ represents the degree of crosstalk between the two qubits. The Hamiltonian term $H_2$ can result from microwave line crosstalk on Q2's line when explicitly driving down Q1's line, as stated above, or from directly driving down Q2's line at $f_1$; the form of the Hamiltonian is the same. The goal of our crosstalk compensation procedure is to null the conditional Rabi drive term, $(X + Y) \otimes Z$, that is produced through Q2 when we drive down Q1's line. We do so by adding an additional single-qubit compensation pulse (with equal amplitude but opposite phase) down Q2's line that cancels this term, thus nulling the field at Q2. This method can be done pairwise for any nearest-neighbors, and operates on the assumption that the crosstalk is sufficiently local to a few qubits. For non-nearest neighbors, we use a similar method that is instead based upon minimization of unwanted AC Stark shifts from crosstalk measured via Ramsey spectroscopy \cite{rudinger2021experimental}. The benchmarking fidelities reported in the text and supplement reflect gates tuned with this compensation.

The two-qubit CNOT gates are generated utilizing the cross-resonance \cite{paraoanu2006microwave, rigetti2010fully, chow2011simple, sheldon2016procedure} effect. Each CNOT gate is applied serially with respect to all other gates. However, neighboring spectator qubits still suffer from conditional and unconditional phase errors due to crosstalk. Unconditional errors are corrected with a virtual phase gate and conditional errors are corrected with a refocusing pulse on the spectator qubit. These methods are further discussed in Section \ref{sec_cb_cer}.

\section{Randomized Benchmarking}\label{rb}
\noindent
We perform isolated and simultaneous single-qubit randomized benchmarking \cite{emerson2005scalable, knill2008randomized, dankert2009exact, magesan2011scalable} (RB) and isolated two-qubit RB to measure the infidelity of our single- and two-qubit gates. Furthermore, we perform unitary RB \cite{wallman2015estimating, dirkse2019efficient} to measure the infidelity due to coherent errors, giving us a quantitative measure of the fraction of the total error rate due to coherent errors versus stochastic errors. All error rates in Tables \ref{tab:sqp} and \ref{tab:tqp} are defined in terms of the process infidelity $e_F(\mathcal{E})$, which is equivalent to the average gate infidelity $r(\mathcal{E})$ (defined in Eq.~\ref{motivation}.\ref{ave_gate_inf}) up to the dimensionality constant $d = 2^n$ for $n$ qubits:
\begin{equation}\label{process_inf}
    e_F(\mathcal{E}) = r(\mathcal{E})\frac{d+1}{d}.
\end{equation}
We note that the process infidelity can be used to compute composite error rates for tensor products of processes, independent of the dimension $d$ of each process, whereas the average gate infidelity cannot.

\section{Cycle Benchmarking and Cycle Error Reconstruction}\label{sec_cb_cer}

\begin{figure}
\centerline{\includegraphics[width=0.8\textwidth]{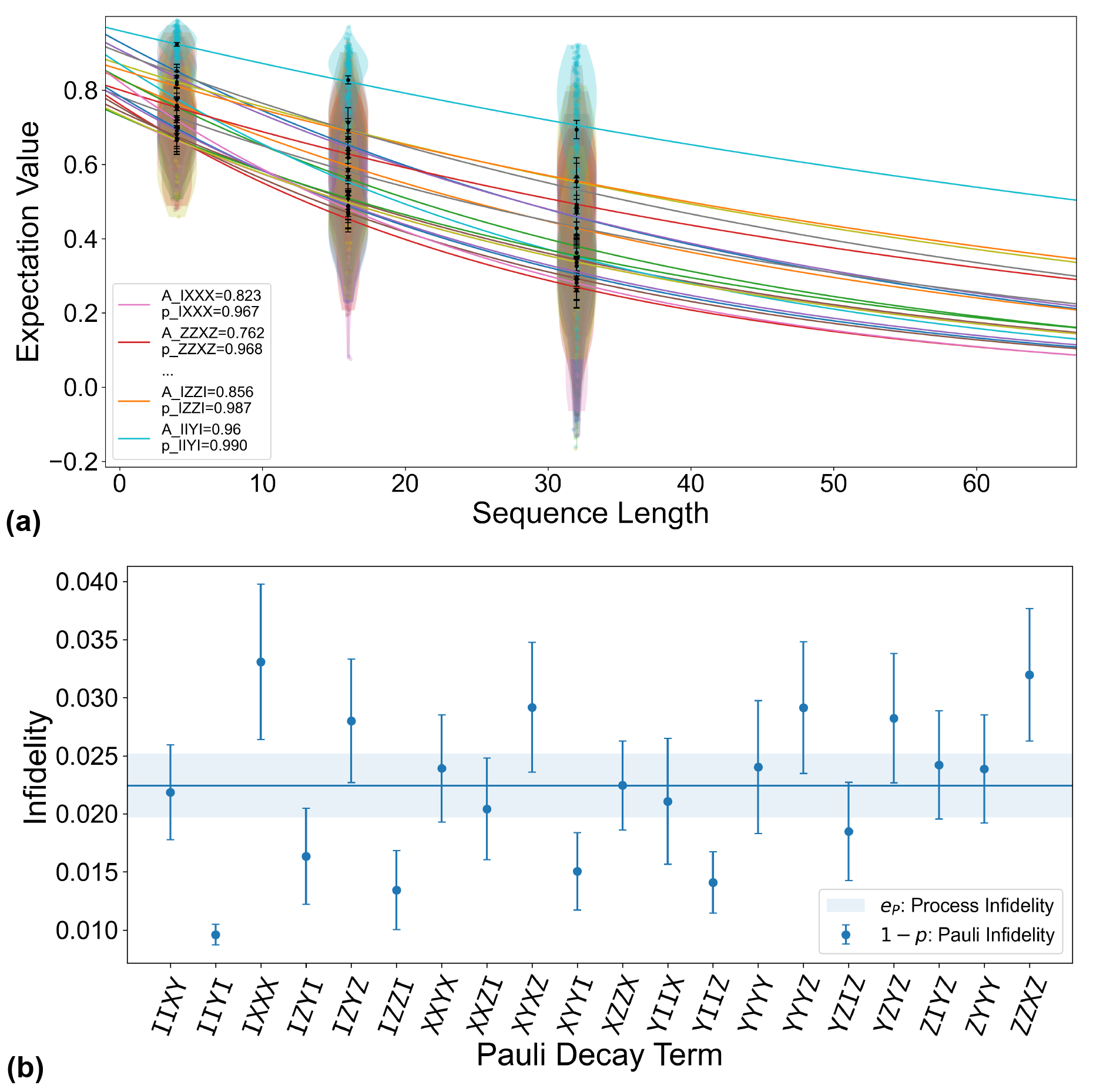}}
\caption{
    \textbf{Cycle benchmarking results on the four-qubit cycle containing only identity gates.
    \textbf{(a)} \normalfont{All measured Pauli decays as a function of sequence length. An exponential decay of the form $Ap^m$ can be fit to the cycle of interest as a function of circuit depth for each basis preparation state; these are referred to as Pauli decays.}
    \textbf{(b)} \normalfont{Comparison of the infidelities of each Pauli decay measured by CB. The total process infidelity is equal to 2.2 $\times$ 10$^{\text{-2}}$ (1.4 $\times$ 10$^{\text{-3}}$). This value represents the average process infidelity of any arbitrary cycle of parallel Pauli gates on all four qubits.}}}
    \label{fig:Figure_S_CB_identity_cycle}
\end{figure}

\begin{figure}
\centerline{\includegraphics[width=0.8\textwidth]{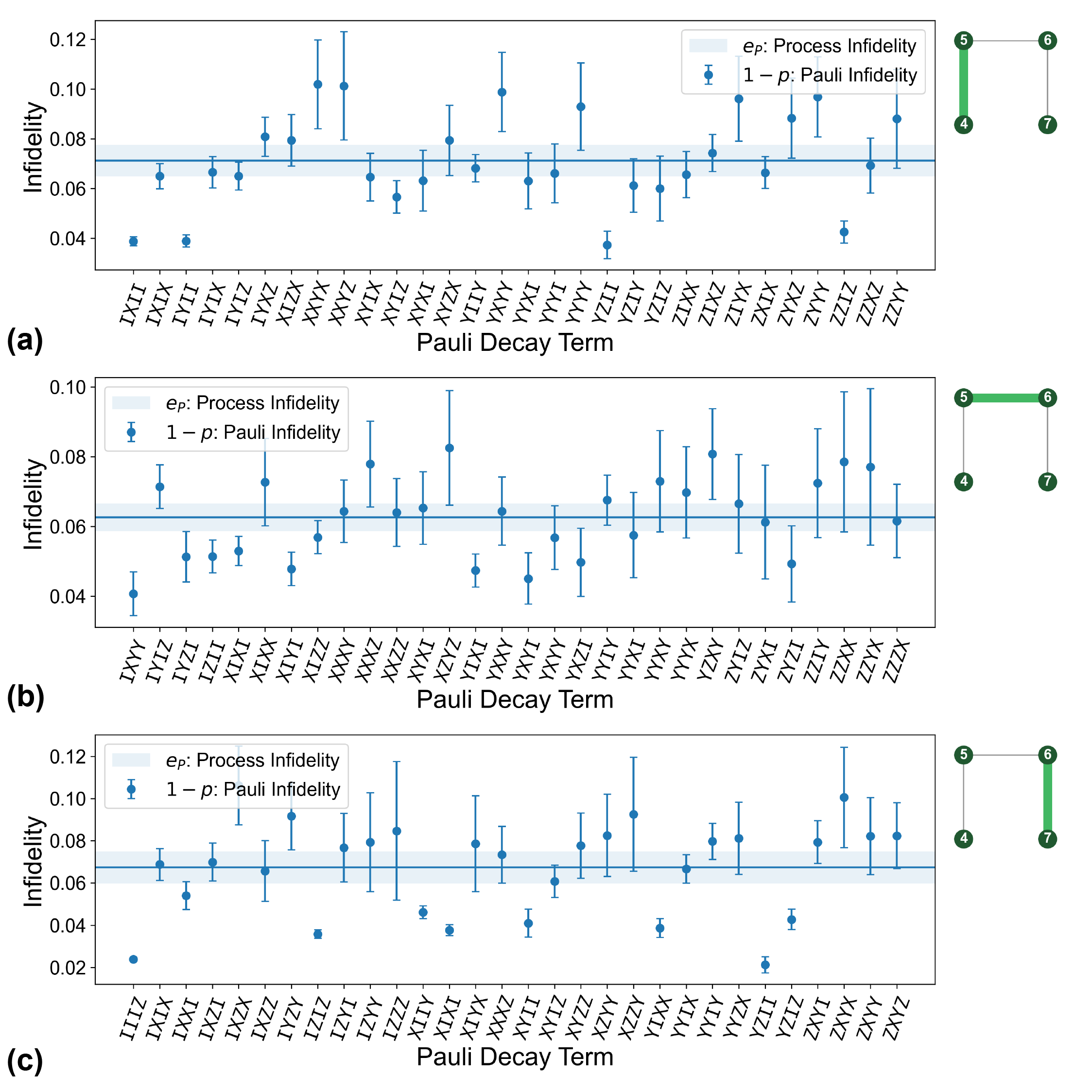}}
\caption{
    \textbf{Cycle benchmarking results for four-qubit cycles containing a single CNOT gate and identity gates on the spectator qubits.
    \textbf{(a)} \normalfont{CNOT between qubits 5 (control) and 4 (target), $e_F = 7.1 \times 10^{\text{-2}} \; (3.2 \times 10^{\text{-3}})$.}
    \textbf{(b)} \normalfont{CNOT between qubits 6 (control) and 5 (target), $e_F = 6.3 \times 10^{\text{-2}} \; (2.0 \times 10^{\text{-3}})$.}
    \textbf{(c)} \normalfont{CNOT between qubits 7 (control) and 6 (target), $e_F = 6.7 \times 10^{\text{-2}} \; (3.9 \times 10^{\text{-3}})$.}}}
    \label{fig:Figure_S_CB_CNOT_cycles}
\end{figure}

\begin{figure}
\centerline{\includegraphics[width=0.8\textwidth]{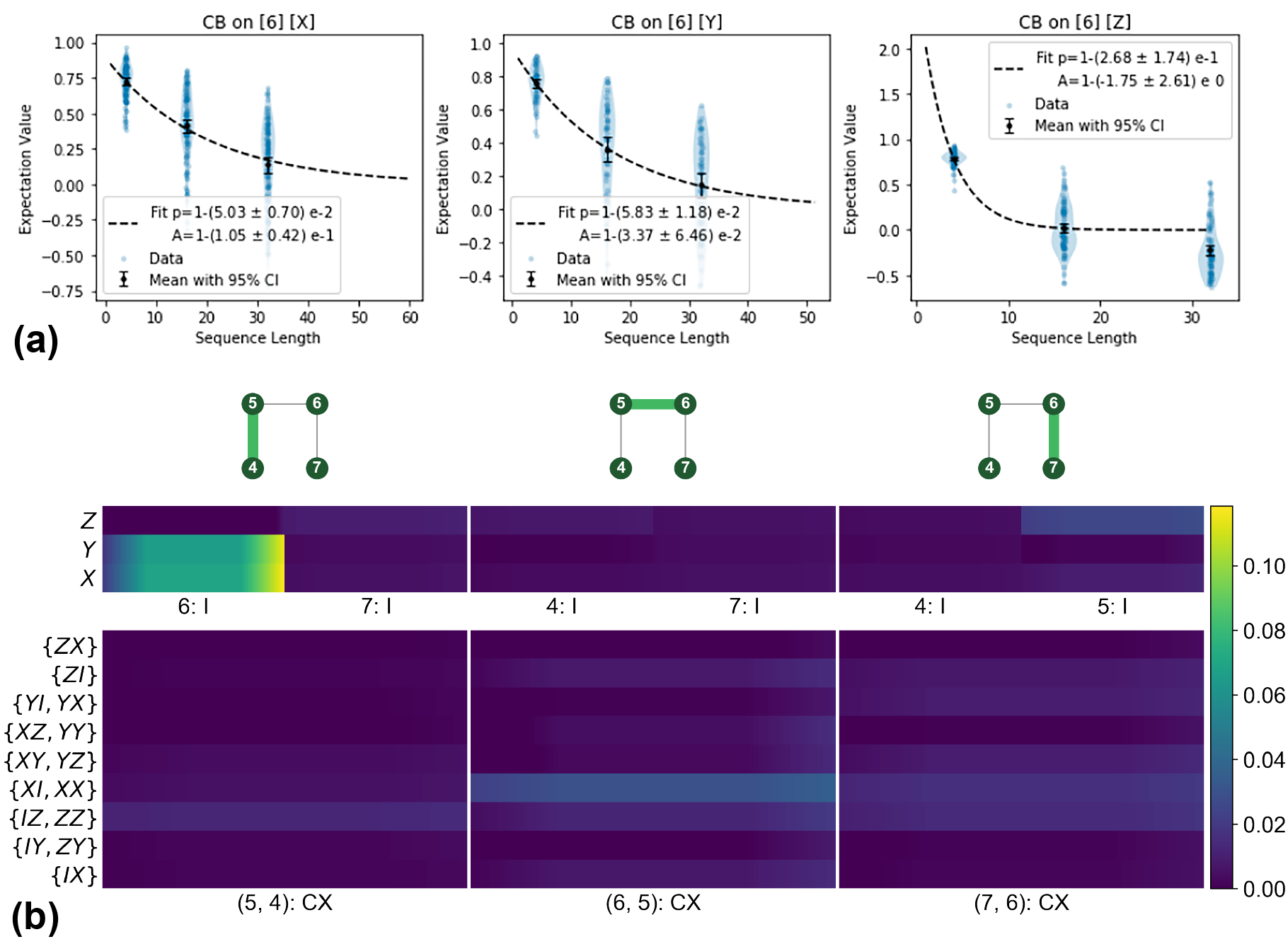}}
\caption{
    \textbf{Cycle benchmarking and cycle error reconstruction results before any targeted gate tuneup.
    \textbf{(a)} \normalfont{CB results marginalized over qubit 6 showing the different rates of decay for $X$, $Y$, and $Z$ operators during the CNOT between qubits 4 and 5. The sharp decay in $Z$ suggests that the dominant errors are due to non-commuting Pauli operators (e.g., $X$ or $Y$).}
    \textbf{(b)} \normalfont{CER results show that the most dominant Pauli errors for all multi-qubit cycles in our system were due to both $X$ and $Y$ errors on qubit 6 during the CNOT between qubits 4 and 5, before any targeted gate tuneup was performed. This information was used to correct for the errors on qubit 6 using a targeted refocusing pulse unconditional on the state of qubit 7, leading to a $\sim 4 \times$ reduction in the largest error rates (see Fig.~\ref{fig:Figure_S_KNR}).}}}
    \label{fig:Figure_S_CB_KNR_first}
\end{figure}

\begin{figure}
\centerline{\includegraphics[width=0.8\textwidth]{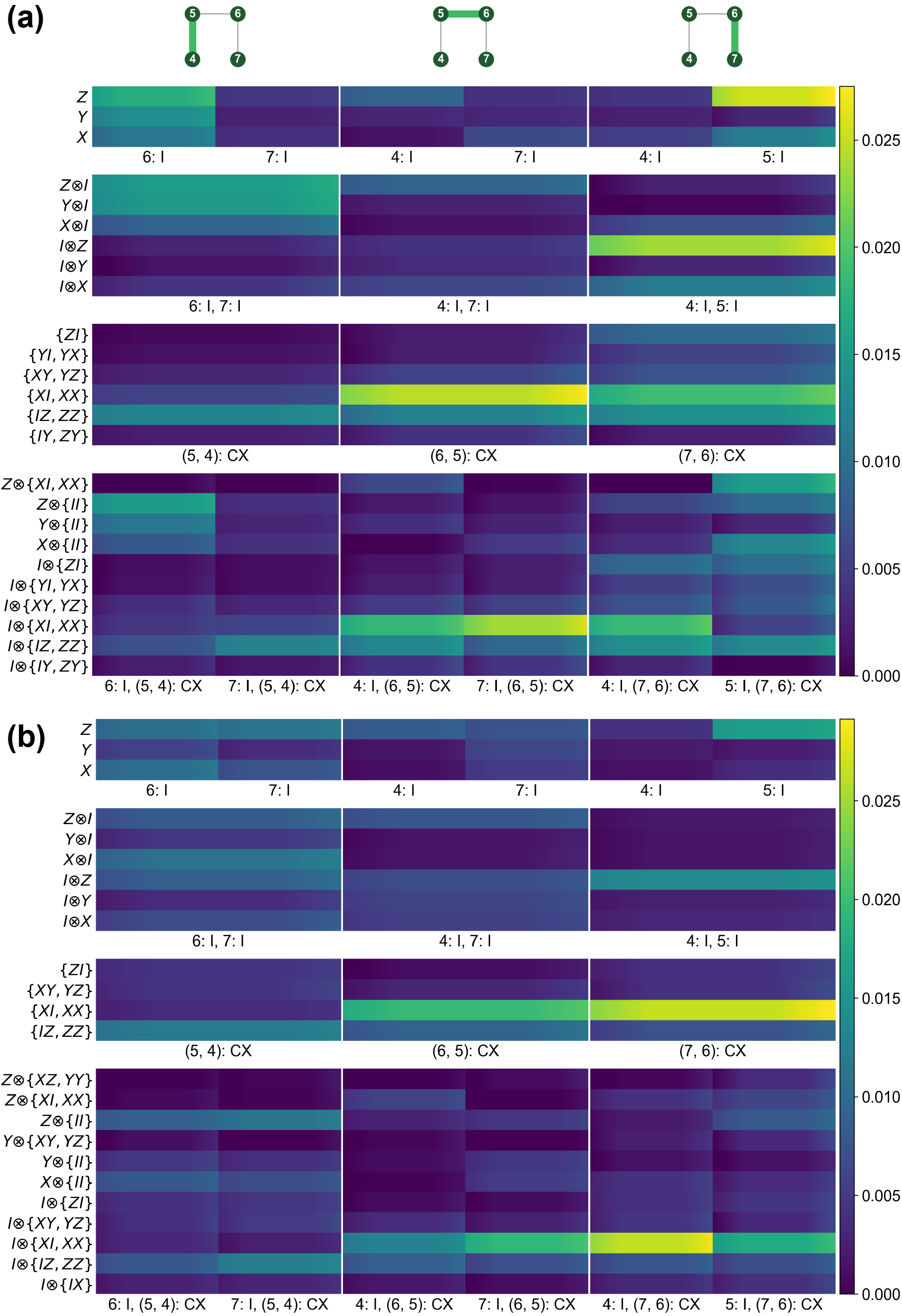}}
\caption{
    \normalfont{Cycle error reconstruction results measured before the (\textbf{a}) random circuits of variable depth and (\textbf{b}) quantum Fourier transform experiments. Rows in which all errors are below 20\% of the maximum value have been omitted for clarity. The largest error rates are $\sim 4 \times$ lower than the initial CER data shown in Fig.~\ref{fig:Figure_S_CB_KNR_first} due to the targeted gate tuneup.}}
    \label{fig:Figure_S_KNR}
\end{figure}

\noindent
Cycle benchmarking \cite{erhard2019characterizing} (CB) is a scalable benchmarking protocol for measuring the local and global errors impacting multi-qubit gate cycles. Under CB, the process fidelity of any parallel gate cycle can be measured by preparing the system in a Pauli basis state, interleaving the cycle of interest between cycles of random $n$-qubit Pauli operators randomly sampled from the full $P^{\otimes n}$ Pauli group, and measuring the error rate as a function of sequence depth. For each basis preparation, an exponential decay of the form $Ap^m$ can be fit to the dressed cycle (composition of the cycle of interest with the random Pauli cycle), where $A$ is the SPAM parameter, $p$ is the decay parameter, and $m$ is the sequence length. Therefore, the exponential decays in CB are often referred to as ``Pauli decays'' and are labeled by the basis preparation state. CB effectively measures the eigenvalues $p$ in the Pauli transfer matrix (PTM) superoperator formalism. For example, CB performed on a two-qubit cycle would produce a PTM with the following diagonal components,
\begin{equation*}
    \Lambda = \begin{psmallmatrix}
                    1 &  &  &  &  &  &  &  &  &  &  &  &  &  &  &  \\
                      & p^m_{IX} &  &  &  &  &  &  &  &  &  &  &  &  &  &  \\
                      &  & p^m_{IY} &  &  &  &  &  &  &  &  &  &  &  &  &  \\
                      &  &  & p^m_{IZ} &  &  &  &  &  &  &  &  &  &  &  &  \\
                      &  &  &  & p^m_{XI} &  &  &  &  &  &  &  &  &  &  &  \\
                      &  &  &  &  & p^m_{XX} &  &  &  &  &  &  &  &  &  &  \\
                      &  &  &  &  &  & p^m_{XY} &  &  &  &  &  &  &  &  &  \\
                      &  &  &  &  &  &  & p^m_{XZ} &  &  &  &  &  &  &  &  \\
                      &  &  &  &  &  &  &  & p^m_{YI} &  &  &  &  &  &  &  \\
                      &  &  &  &  &  &  &  &  & p^m_{YX} &  &  &  &  &  &  \\
                      &  &  &  &  &  &  &  &  &  & p^m_{YY} &  &  &  &  &  \\
                      &  &  &  &  &  &  &  &  &  &  & p^m_{YZ} &  &  &  &  \\
                      &  &  &  &  &  &  &  &  &  &  &  & p^m_{ZI} &  &  &  \\
                      &  &  &  &  &  &  &  &  &  &  &  &  & p^m_{ZX} &  &  \\
                      &  &  &  &  &  &  &  &  &  &  &  &  &  & p^m_{ZY} &  \\
                      &  &  &  &  &  &  &  &  &  &  &  &  &  &  & p^m_{ZZ}
                  \end{psmallmatrix},
\end{equation*}
where $p^m_P$ is the eigenvalue of the Pauli basis $P$ at a sequence depth of $m$. In the PTM formalism, the total process fidelity is the weighted average of the diagonal components,
\begin{equation}
    F = \frac{1}{4^n} \sum_{P \in \mathbb{P}^{\otimes n}} \Lambda_{P,P},
\end{equation}
where $\mathbb{P}^{\otimes n} = \{I, X, Y, Z\}^{\otimes n}$ the set of $4^{n}$ generalized Pauli operators. Thus, the process infidelity is $e_F = 1 - F$. However, it is worth noting that, in general, the process infidelity of any dressed cycle also includes the errors due to the twirling operators,
\begin{equation}\label{cb_process_infidelity_CT}
    e_F = 1 - \mathbb{E}\left[ \operatorname{Tr} \left(
    \mathcal{T}^\dagger \mathcal{C}^\dagger \tilde{\mathcal{C}} \tilde{\mathcal{T}} \right)\right],
\end{equation}
where $\mathcal{T}$ is the superoperator of the cycle of Pauli operators, $\mathcal{C}$ is the superoperator of the cycle of interest, a tilde denotes the experimental (noisy) implementation of each superoperator, and $\mathbb{E}$ is the weighted average over all randomly-chosen $n$-qubit Pauli operators. In the limit of perfect Pauli twirling, the above equation simplifies to the process infidelity of the cycle of interest alone:
\begin{equation}\label{cb_process_infidelity_C}
    e_F = 1 - \operatorname{Tr} \left(\mathcal{C}^\dagger \tilde{\mathcal{C}} \right).
\end{equation}
The effective noise under CB is a Pauli channel which, in the Kraus representation, maps any $n$-qubit density matrix $\rho$ into
\begin{equation}\label{pauli_channel}
    \mathcal{E}(\rho) = \sum_{P \in \mathbb{P}^{\otimes n}} c_P P \rho P^\dagger,
\end{equation}
with $c_P$ the relative probability of an error due to $P$. This is the same as the tailored noise under randomized compiling \cite{wallman2016noise} (RC), since both CB and RC utilize Pauli twirling. Therefore, CB gives accurate estimates of the process infidelity of a cycle when performed in an algorithm under RC. We performed CB measurements on a four-qubit cycle containing identity gates in order to benchmark the average process infidelity of our gate cycles containing only single-qubit gates (i.e., easy gate cycles; see Fig.~\ref{fig:Figure_S_CB_identity_cycle}), as well as four-qubit cycles containing a single CNOT gate and identity gates on the other two spectator qubits (i.e., hard gate cycles; see Fig.~\ref{fig:Figure_S_CB_CNOT_cycles}).

Cycle error reconstruction \cite{erhard2019characterizing, flammia2019efficient} (CER) results are based on targeted CB measurements in which specific Pauli decays are chosen to estimate the error rates afflicting subsets of the gates or idle qubits in the specific cycle of interest. The Pauli error rates estimated by CER are the coefficients $c_P$ in the Kraus representation of a Pauli channel in Eq.\ref{sec_cb_cer}.~\ref{pauli_channel}. One can convert the between the Pauli eigenvalues $p_P$ and the Kraus coefficients $c_P$ using a linear transformation,
\begin{equation}
    p_P = \mathbb{W} c_P,
\end{equation}
where $\mathbb{W}$ is an $n$-qubit Walsh-Hadamard transform (see Ref.~\cite{flammia2019efficient}, Section B for further details). The Pauli decays in CB are dual to the Pauli operators which cause errors, so to measure the error rate $c_P$ of some fixed Pauli $P$, we measure a set Pauli decays that commute and anti-commute with $P$, and then use this info to reconstruct the error of $P$ via linear inversion: $c_P = \mathbb{W}^{-1} p_P$. An example of this duality of Pauli decays to Pauli errors is given in Fig.~\ref{fig:Figure_S_CB_KNR_first}.

CER measurements were performed throughout the duration of this experiment. Here, we present the results measured before the random circuits of variable depth and quantum Fourier transform experiments, as shown in Fig.~\ref{fig:Figure_S_KNR}. Due to drift and modifications in the tuneup parameters, these plots are not identical. The results are based on measurements of each hard gate cycle involving a CNOT gate on one pair of qubits and identity gates applied to the spectator qubits, represented by the three different columns. In these plots, the y-axis indicates the type of error, the x-axis indicates the qubit(s) on which the error occurs, and the color indicates the marginal error rate of each type of error occurring from all Pauli contributions, with the color gradient across each cell defining the 95\% confidence interval. For example, a marginal $X$ error on qubit 4 is the total summed probability of all Pauli errors that act as $X$ on qubit 4 (e.g., $\mathcal{P}(XIII) + \mathcal{P}(XYZI) + \mathcal{P}(XXYZ) + ...$); in particular, it is not just the probability of $XIII$ alone. The first row of subplots shows single-body gate errors on idle qubits; the second row of subplots shows correlated gate errors between idling qubits (e.g., $Z \otimes I$ is the probability of a $Z$ error on the first qubit and no error on the second qubit, $Z \otimes Z$ is the $ZZ$ coupling between qubits, etc.); the third row of subplots shows single-body gate errors on qubits involved in a CNOT gate, where curly brackets indicate errors that cannot be distinguished due to degeneracies; and the last row of subplots shows correlated errors between an idle qubit and a CNOT on a different pair of qubits. Here, a $k$-body error refers to an error correlator that contains $k$ non-identity Pauli operators. We only measured 1- and 2-body errors, as the most dominant error syndromes occur at this level. The tensor notation $\otimes$ between $k$-body errors indicates correlators between product states, whereas the lack of tensors indicates correlators for entangled qubits. As seen in Fig.~\ref{fig:Figure_S_KNR}, the most dominant errors in our system are 1-body errors (up to any potential degeneracy with 2-body terms). We also note that, in general, it is possible to reconstruct all 1-body marginals with $\mathcal{O}(1)$ Pauli decays and all 2-body marginals with $\mathcal{O}(\log(n))$ Pauli decays for $n$ qubits. However, while some $k$-body errors can efficiently be estimated for $k > 2$, it becomes exponentially expensive to estimate \textit{all} $n$-body errors.

As outlined in Fig. 2a of the main text, CER results can be utilized for targeted gate tuneup. Initial CER measurements performed during the experiment (Fig.~\ref{fig:Figure_S_CB_KNR_first}) highlighted that the most dominant errors in our system were $X$ and $Y$ errors on qubit 6 during the CNOT between qubits 4 and 5, occurring at a marginal error rate of $\sim$0.10. Using this knowledge, we compensated for these spectator errors using a refocusing pulse on qubit 6, unconditional on the state of qubit 7. Other common errors encountered during our targeted gate tuneup were $Z$ errors on spectator qubits during CNOT gates. $Z$ errors on spectator qubits (e.g., qubit 7) adjacent to control qubits (e.g., qubit 6 in the (6, 5) CNOT) were most dominantly caused by off-resonantly driving the spectator qubit via crosstalk from the control qubit pulse lines, causing an unwanted cross-resonance (CR) effect. To address these errors, slow $R_x(2\pi)$ pulses were added to the spectator qubit to echo away the accumulated phase error. $Z$ errors on spectator qubits (e.g., qubit 4) adjacent to target qubits (e.g., qubit 5 in the (6, 5) CNOT) were corrected using virtual $Z$ gates, as these errors were not due to entangling effects.

The CER results in Fig.~\ref{fig:Figure_S_KNR} highlight the remaining errors in our system before the random circuits of variable depth and quantum Fourier transform experiments. Compared to the initial results in Fig.~\ref{fig:Figure_S_CB_KNR_first}, these error syndromes are much more broadly distributed, making further targeted corrections less fruitful and/or difficult to implement without causing unwanted errors on other error channels as a result. Several features stand out from these results: (1) there remains a large residual $Z$ error on qubit 5 during the (7, 6) CNOT; (2) error correlators of the form $\{XI, XX\}$ remain dominant on the CNOT qubit pairs, with the exception of (5, 4). For (1), it is unknown why the methods described above for correcting $Z$ errors on spectator qubits do not fully correct for these errors, although it is worth mentioning that these errors were less prominent in the data taken before the quantum Fourier transform experiments. While the physical origin of (2) is not fully known, there are indicators that suggest that it is due to a CR-type interaction between a spectator qubit (e.g., qubit 4 or 5) and the adjacent target qubit (e.g. qubit 5 or 6) in a CNOT, with the spectator qubit acting as the control qubit in the unwanted CR interaction; this effect would be consistent with the spectator qubit driving an $X$ over-rotation error on the adjacent target qubit. Tellingly, no such dominant error is seen for the (5, 4) CNOT pair, since qubit 4 is not adjacent to any actively-used spectator qubit on the processor. Finally, while our qubit architecture lends itself to always-on $ZZ$ coupling between nearest-neighbor qubits, we see that this is not a dominant residual term in Fig.~\ref{fig:Figure_S_KNR}. We believe this due to some level of dynamical decoupling occurring from Pauli twirling during CB, and from dynamical decoupling on target gates due to the $ZX$-type interaction of the CR gate.

\section{Quadratic Impact of Coherent Errors}\label{sec_coherent_error}
\noindent
Coherent errors acting on a single qubit can be modeled as a unitary rotation operator,
\begin{equation}\label{coherent_error}
    U(\mathbf{\hat{n}}, \theta) = e^{-i\theta\mathbf{\hat{n} \cdot \boldsymbol\sigma}/2},
\end{equation}
where $\theta$ is the rotation angle relative to the intended target state, $\mathbf{\hat{n}}$ is the axis of rotation, and $\boldsymbol\sigma$ the Pauli vector. Coherent errors map pure states to pure states, $\rho = \ket{\psi}\bra{\psi} \mapsto \mathcal{E}(\rho) = U \ket{\psi}\bra{\psi} U^\dagger$, and therefore do not result in decoherence.

To understand how coherent errors impact the average gate fidelity versus norm-based metrics for error, even in the single-step limit and not just through the potential for adversarial accumulation of errors under compositions, consider an error associated with a rotation about the $z$-axis \textit{for a single computational gate}, with evolution operator
\begin{align}
    U(z, \theta) &= \exp\left( - i \frac{\theta}{2} \sigma_z \right) \nonumber \\
    &= \cos(\theta/2) - i\sigma_z\sin(\theta/2) \nonumber \\
    &= \left( \begin{array}{cc} e^{-i\theta/2} & 0 \\ 
        0 & e^{i\theta/2} \end{array} \right).
\end{align}
The PTM superoperator for this coherent error for a single computational gate takes the form
\begin{equation}
   \Lambda = \left(
        \begin{array}{cccc}
            1 & 0 & 0 & 0 \\
            0 & \cos(\theta) & - \sin(\theta) & 0 \\
            0 & \sin(\theta) & \cos(\theta) & 0 \\
            0 & 0 & 0 & 1 \\
        \end{array}
        \right).
\end{equation}
In the small error limit, we see that the diagonal terms have matrix elements proportional to $\cos (\theta) \sim 1 - \theta^2$, and thus have errors proportional to $\theta^2$, whereas the off-diagonal terms appear at the scale $\sin (\theta ) \sim \theta$. Under RC, the off-diagonal terms are suppressed, and under \textit{a perfectly implemented} twirl this leads to
\begin{equation}
   \Lambda = \left(
        \begin{array}{cccc}
            1 & 0 & 0 & 0 \\
            0 & \cos(\theta) & 0 & 0 \\
            0 & 0 & \cos(\theta) & 0 \\
            0 & 0 & 0 & 1 \\
        \end{array}
        \right).
\end{equation}
This corresponds to the general error reduction from $\sqrt{r(\mathcal{E})} \simeq \theta$ to $r(\mathcal{E}) \simeq \theta^2$ that is achieved by RC, where $r(\mathcal{E})$ is the average error rate.

We also observe that error assessments based on the average gate infidelity or process infidelity are only sensitive to the diagonal terms of the error process, and thus cannot benefit from the suppression of the off-diagonal terms through RC. In contrast, norm-based error-metrics, such as the total variation distance (TVD) $d_\textrm{TV}(\mathcal{P}, \mathcal{P}_\textrm{ideal})$ and the diamond distance $\epsilon_\diamond(\mathcal{E} - I)$ (both defined below), generally will be sensitive to the off-diagonal terms in the error process, and thus generally benefit from RC.

However, it should be noted that, whereas the diamond distance is always sensitive to the off-diagonal terms,  the TVD is a basis-dependent metric and is only sensitive to the off-diagonal terms if the (ideal) target state is coherently spread across the measurement basis, where a larger coherent spread implies a greater number of off-diagonal terms that can contribute to the error metric. This observation is consistent with the data shown in Fig.~\ref{fig:Figure_S_QFT_2_3_qubits}, where the degree of uniformity of the target state across the measurement basis is found to correlate with the degree of error suppression that is observed experimentally. We discuss this more in depth below, in Section \ref{sec_tvd_fidelity}.

\section{Motivation for Noise Tailoring via Randomized Compiling}\label{motivation}
\noindent
Here, we outline some theoretical motivation for why converting coherent errors into stochastic noise provides significant performance improvements for both NISQ applications and fault-tolerant quantum error correction. First, consider three common methods for assessing error rates on a gate or a cycle of parallel gates: (i) the error rate $r(\mathcal{E})$, defined as the average gate infidelity of a noise process $\mathcal{E}$ relative to the identity operation $I$, which is the standard quantity measured by RB (and is equivalent to the process infidelity $e_F$, defined in Eq.~\ref{rb}.\ref{process_inf}); (ii) the diamond distance $\epsilon_\diamond(\mathcal{E} - I)$ often used to compute fault-tolerant thresholds \cite{aharonov2008fault}; and (iii) the total variation distance (TVD) $d_\textrm{TV}(\mathcal{P}, \mathcal{P}_\textrm{ideal})$ between the ideal $\mathcal{P}_\textrm{ideal}$ and experimental $\mathcal{P}$ probability distributions, which are induced by the noiseless $I$ and noisy $\mathcal{E}$ error models, respectively. The infidelity $r(\mathcal{E})$ is defined via the average gate fidelity of a noisy operation $\mathcal{E}(\rho)$,
\begin{equation}\label{ave_gate_inf}
    r(\mathcal{E}) = 1 - \int d\psi \bra{\psi}\mathcal{E}(\ket{\psi}\bra{\psi})\ket{\psi},
\end{equation}
and $\epsilon_\diamond(\mathcal{E}-I)$ is defined as the diamond distance from the identity operator \cite{kitaev1997quantum, kueng2016comparing},
\begin{equation}\label{diamond_norm}
    \epsilon_\diamond(\mathcal{E}-I) = \frac{1}{2} \big|\big| \mathcal{E} - I \big|\big|_\diamond = \frac{1}{2} \max_{\rho} \big|\big| \big[ I \otimes \mathcal{E} - I \otimes I \big](\rho) \big|\big|_1,
\end{equation}
where the maximum is taken over all pure states and $\big|\big| X \big|\big|_1 = \textrm{Tr}\sqrt{X^\dagger X}$. 
Finally, the TVD measures the statistical distance between two probability distributions over a set of possible outcomes:
\begin{equation}\label{total_variation_distance}
    d_\textrm{TV}(\mathcal{P}, \mathcal{P}_\textrm{ideal}) = \frac{1}{2} \sum_{x \in X} |\mathcal{P}(x) - \mathcal{P}_\textrm{ideal}(x)|,
\end{equation}
where $\mathcal{P}_\textrm{ideal}(x)$ is the ideal probability of measuring a bit string $x$ in a set of possible bit strings $X$, and $\mathcal{P}(x)$ is the measured (noisy) probability. The TVD is a basis-dependent metric which determines the probability of obtaining an incorrect solution (i.e., distribution of bit strings) under a NISQ application, and is bounded between 0 and 1, with 0 (1) indicating that the correct solution is always (never) measured.

These common error metrics are related by the following inequalities in $d = 2^n$ dimensions,
\begin{equation}\label{bounds1}
    r(\mathcal{E})\frac{d+1}{d} \leq \epsilon_\diamond(\mathcal{E - I}) \leq \sqrt{r(\mathcal{E})}\sqrt{d(d+1)},
\end{equation}
and
\begin{equation}\label{bounds2}
d_\textrm{TV}(\mathcal{P}, \mathcal{P}_\textrm{ideal}) \leq \epsilon_\diamond(\mathcal{E - I}).
\end{equation}
The upper bound in Eq.~\ref{bounds2} follows from considering the definition of 
$\epsilon_\diamond(\mathcal{E}-I)$, when defining  $\rho_\mathcal{E} = I \otimes \mathcal{E} (\rho)$ and $\rho_\mathcal{I} = I \otimes I (\rho)$, so that
\begin{equation}
    \frac{1}{2} || \rho_\mathcal{E} - \rho_\mathcal{I}||_1 \leq   \epsilon_\diamond(\mathcal{E}-I) .
\end{equation}
Then, defining the partial trace taking any $\rho$ (on the dilated Hilbert space) to $\sigma$, a density operator on the (reduced) base Hilbert space, and noting that the partial trace is a completely positive trace-preserving (CPTP) map and that the norm $|| \cdot ||_1$ is contractive under CPTP maps, it follows that
\begin{equation}
    || \sigma_\mathcal{E} - \sigma_\mathcal{I}||_1  \leq  || \rho_\mathcal{E} - \rho_\mathcal{I}||_1. 
\end{equation}
Finally, if we define $\mathcal{P}_\textrm{ideal}$ ($\mathcal{P}$) as the probability distribution induced by the measurement of $\sigma_\mathcal{I}$ ($\sigma_\mathcal{E}$), then 
the definition of the TVD implies
\begin{equation}
d_\textrm{TV}(\mathcal{P}, \mathcal{P}_\textrm{ideal}) \leq \frac{1}{2} || \sigma_\mathcal{E} - \sigma_\mathcal{I}||_1 \leq \epsilon_\diamond(\mathcal{E - I}).
\end{equation}

While $\epsilon_\diamond(\mathcal{E - I})$ is generally difficult to measure experimentally, any stochastic error model $\mathcal{E}_{\textrm{RC}}$ (e.g., achieved through RC) is known \cite{magesan2012characterizing} to saturate the lower bound in Eq.~\ref{bounds1}, which implies
\begin{equation}
d_\textrm{TV}(\mathcal{P}_{\textrm{RC}}, \mathcal{P}_\textrm{ideal}) \leq  \epsilon_\diamond(\mathcal{E_{\textrm{RC}} - I}) =  r(\mathcal{E_{\textrm{RC}}})\frac{d+1}{d}.
\end{equation}
In contrast, for general coherent errors, $\epsilon_\diamond(\mathcal{E - I})$ is known to saturate the upper bound of Eq.~\ref{bounds1}, scaling with $\sqrt{r(\mathcal{E})}$ \cite{wallman2014randomized, sanders2015bounding}. For typical two-qubit RB error rates of $r(\mathcal{E}) \simeq 10^{-2}$ in today's NISQ systems, coherent errors at this scale can contribute as much as $\sqrt{r(\mathcal{E})} \simeq 10^{-1}$ to the probability of an incorrect outcome for each two-qubit gate in the circuit, whereas for stochastic errors the impact remains at the $10^{-2}$ scale, and can be directly compared to fault-tolerant thresholds based on the diamond norm. As gate error rates scale down, the quadratic reduction $\sqrt{r(\mathcal{E})} \rightarrow r(\mathcal{E})$ from RC can provide significantly greater performance gains by simply tailoring residual coherent errors into stochastic Pauli noise at the compiler level. This quadratic reduction in the worst-case error rate enables the direct comparison of benchmarked gate fidelities with fault-tolerance thresholds for Pauli errors, making fault-tolerant quantum computation possible with gate fidelities comparable to those realized in modern-day experiments.

\section{TVD vs. Fidelity: Impact of RC on Linear vs. Non-Linear Error Metrics}\label{sec_tvd_fidelity}
\noindent
The primary benefit of RC in the many-randomization limit results in a quadratic improvement (i.e., $\sqrt{r(\mathcal{E})} \longrightarrow r(\mathcal{E})$) for norm-based error metrics, such as the diamond norm (Eq.~\ref{motivation}.\ref{diamond_norm}) and the TVD (Eq.~\ref{motivation}.\ref{total_variation_distance}). However, this quadratic improvement depends on the linearity of the error metric and the degree to which it is sensitive to off-diagonal terms in the error process. Here, we compare the difference between the fidelity of a quantum state $\rho$ with the ideal (pure) quantum state $\rho_\textrm{ideal} = \ketbra{\phi}{\phi}$, 
\begin{equation}
    \mathcal{F} = \Tr[\rho \ketbra{\phi}{\phi}] = \bra{\phi} \rho \ket{\phi},
\end{equation}
and the total variation distance (TVD). Whereas the fidelity is always insensitive to off-diagonal terms in the PTM error process, the TVD is generally sensitive to these terms, except for the case in which the target state is an eigenstate of the measurement basis. We explore this in detail below.

Both of these metrics have physical motivations: the fidelity is well-understood in quantum mechanics to represent the inner product (or overlap) between two quantum states. In the case of the TVD, it has both a statistical and physical interpretation. Statistically, the TVD quantifies the maximum probability that measuring the same event $E$ (e.g., a specific bit string or a set of bit strings) can be distinguished between the two distributions $\mathcal{P}$ and $\mathcal{P}_\textrm{ideal}$,
\begin{equation}
d_\textrm{TV}(\mathcal{P}, \mathcal{P}_\textrm{ideal}) = \max_E |\mathcal{P}(E) - \mathcal{P}_\textrm{ideal}(E)| = \max_E \Big| \sum_{x \in E}(\mathcal{P}(x) - \mathcal{P}_\textrm{ideal}(x)) \Big|,
\end{equation}
where the maximization is taken over all possible subsets $E$ from the full set of bit strings $X$. Given this definition, we adopt the convenient interpretation that the TVD represents the \textit{probability of measuring the incorrect solution} to a quantum algorithm. For a countable set (as is the case when measuring bit strings), the TVD is related to the L1-norm through the following identity:
\begin{equation}\label{tvd_l1_norm}
    d_\textrm{TV}(\mathcal{P}, \mathcal{P}_\textrm{ideal}) = \frac{1}{2} \sum_{x \in X} |\mathcal{P}(x) - \mathcal{P}_\textrm{ideal}(x)| = \frac{1}{2} || \mathcal{P} - \mathcal{P}_\textrm{ideal} ||_1.
\end{equation}

The TVD is also closely related trace distance $D(\rho, \rho_\textrm{ideal})$ between $\rho$ and $\rho_\textrm{ideal}$ in quantum mechanics. To see this, we note that $\mathcal{P}(x) = \Tr[\rho M_x]$ and $\mathcal{P}_\textrm{ideal}(x) = \Tr[\rho_\textrm{ideal} M_x]$ can be understood to be the probabilities of measuring a result $x$ given a POVM set $\{ M_x \}$. Thus, the TVD can be written as
\begin{equation}
    d_\textrm{TV}(\mathcal{P}, \mathcal{P}_\textrm{ideal}) = \frac{1}{2} \sum_x \big| \Tr[(\rho - \rho_\textrm{ideal}) M_x] \big|.
\end{equation}
For any quantum states $\rho$ and $\rho_\textrm{ideal}$, it can be shown that the TVD is upper-bounded by the trace distance between $\rho$ and $\rho_\textrm{ideal}$ for any measurement $\{ M_x \}$,
\begin{equation}\label{tvd_trace_distance}
    d_\textrm{TV}(\mathcal{P}, \mathcal{P}_\textrm{ideal}) \le \frac{1}{2} \Tr|\rho - \rho_\textrm{ideal}| \equiv D(\rho, \rho_\textrm{ideal}).
\end{equation}
In the special case in which $\rho$ and $\rho_\textrm{ideal}$ are diagonal in the same basis, $\mathcal{P}(x)$ and $\mathcal{P}_\textrm{ideal}(x)$ can be understood to be the probability distributions induced by measuring the density matrices $\rho$ and $\rho_\textrm{ideal}$ in their common eigenbasis. Writing $\rho$ and $\rho_\textrm{ideal}$ in terms of their spectral decomposition, $\rho = \sum_x \mathcal{P}(x) \ketbra{x}{x}$ and $\rho_\textrm{ideal} = \sum_x \mathcal{P}_\textrm{ideal}(x) \ketbra{x}{x}$, we see that in this case the TVD is exactly equal to the trace distance
\begin{align}
    d_\textrm{TV}(\mathcal{P}, \mathcal{P}_\textrm{ideal}) 
        &= \frac{1}{2} || \mathcal{P} - \mathcal{P}_\textrm{ideal} ||_1 \\
        &= \frac{1}{2} \Tr \Big| \sum_x \big( \mathcal{P}(x) - \mathcal{P}_\textrm{ideal}(x) \big) \ketbra{x}{x} \Big| \\
        &= \frac{1}{2} \Tr | \rho - \rho_\textrm{ideal} | = D(\rho, \rho_\textrm{ideal}),
\end{align}
where we have used Eq.~\ref{tvd_l1_norm} in the first line. More generally, the TVD of probabilities corresponding to a measurement $\{ M_x \}$ will saturate the trace distance in Eq.~\ref{tvd_trace_distance} if $\{ M_x \}$ are projectors on the eigenbasis of $(\rho - \rho_\textrm{ideal})$. Therefore, if $\rho$ and $\rho_\textrm{ideal}$ are close in trace distance, then any measurement of the two density matrices in the same basis will result in probability distributions that are close in total variation distance. Similar to the statistical interpretation of the TVD, the trace distance can interpreted as the maximum probability of distinguishing $\rho$ from $\rho_\textrm{ideal}$ upon measurement,
\begin{equation}
     D(\rho, \rho_\textrm{ideal}) = \max_{\{M_x\}}  d_\textrm{TV}(\mathcal{P}(x), \mathcal{P}_\textrm{ideal}(x)),
\end{equation}
with the maximum is taken over all POVMs $\{ M_x \}$.

Interestingly, the trace distance can be physically understood as being related to the Euclidean distance between $\rho$ and $\rho_\textrm{ideal}$ for the special case of single qubits states on the Bloch sphere. To see this, we can write $\rho$ and $\rho_\textrm{ideal}$ in terms of their respective Bloch vectors $\mathbf{r}$ and $\mathbf{r}_\textrm{ideal}$,
\begin{equation}
    \rho = \frac{1}{2} (I + \mathbf{r} \cdot \boldsymbol{\sigma})
\end{equation}
and
\begin{equation}
    \rho_\textrm{ideal} = \frac{1}{2} (I + \mathbf{r}_\textrm{ideal} \cdot \boldsymbol{\sigma}),
\end{equation}
where $\boldsymbol{\sigma}$ is the single-qubit Pauli vector. The trace distance can be written as
\begin{align}
    D(\rho, \rho_\textrm{ideal}) &= \frac{1}{2} \Tr | \rho - \rho_\textrm{ideal} |  \\
        &= \frac{1}{4} \Tr | (\mathbf{r} - \mathbf{r}_\textrm{ideal}) \cdot \boldsymbol{\sigma} |.
\end{align}
Because the eigenvalues of $\boldsymbol{\sigma}$ are $\pm 1$, the trace of $| (\mathbf{r} - \mathbf{r}_\textrm{ideal}) \cdot \boldsymbol{\sigma} | = 2 | (\mathbf{r} - \mathbf{r}_\textrm{ideal}) |$, and thus
\begin{equation}
    D(\rho, \rho_\textrm{ideal}) = \frac{1}{2} | \mathbf{r} - \mathbf{r}_\textrm{ideal} |.
\end{equation}
In other words, the trace distance between two single-qubit states on the Bloch sphere is exactly equal to one-half the Euclidean distance between their vectors.

Returning to the question of how the utility of RC depends on the linearity of the error metric in the output state $\rho$, we have already shown that the TVD (and thus the trace distance) can greatly benefit from averaging over many randomizations, but the same cannot be said of the fidelity. Consider the noisy output state $\mathcal{E}(\rho)$ prepared by a single randomization under RC. Here, the fidelity is linear in $\rho$:
\begin{equation}
    \mathcal{F} = \bra{\phi} \mathcal{E}(\rho) \ket{\phi},
\end{equation}
where the ideal final state is $\rho_\textrm{ideal} = \ketbra{\phi}{\phi}$. When averaging over many randomizations,
\begin{align}
    \mathcal{F} &= \bra{\phi} \sum_i^N \mathcal{E}_i(\rho) \ket{\phi} \\
        &= \sum_i^N \bra{\phi} \mathcal{E}_i(\rho) \ket{\phi},
\end{align}
where, in the second line, the sum over $N$ randomizations can be taken outside of the inner product. Thus, if a single randomization provides no benefit over the bare circuit, in the many-randomization limit the fidelity is simply an average of $N$ incorrect results.  We note here, however, that a single randomization under RC can still provide an advantage for circuits impacted by structured errors by dynamically-decoupling the qubits from arbitrary non-Markovian noise and breaking up the coherent accumulation of unitary errors (similar to echoed pulse sequences). Therefore, metrics that are linear in $\rho$ can still benefit from RC if the noisy preparation $\mathcal{E}_\textrm{RC}(\rho)$ is more accurate than $\mathcal{E}(\rho)$ in the single-randomization limit.

In contrast, the trace distance (and TVD) is generally not linear in $\rho$, and thus the sum over $N$ randomizations cannot be pulled outside of the absolute value. In fact, the trace distance is convex in its first input, and it follows that 
\begin{align}
     D(\sum_i^N \mathcal{E}_i(\rho), \rho_\textrm{ideal}) 
        & = \frac{1}{2} \Tr \Big| \sum_i^N \mathcal{E}_i(\rho) - \rho_\textrm{ideal} \Big| \\
        & \le \frac{1}{2} \sum_i^N \Tr | \mathcal{E}_i(\rho) - \rho_\textrm{ideal} |.
\end{align}
The equality in the above equation is saturated only for deterministic algorithms, in which the target state is an eigenstate of the measurement basis. To understand why this is the case, consider measurements made in the computation basis $\ket{x} \in \{ \ket{0}, \ket{1} \}$, where $\mathcal{P}(x) = \Tr[\mathcal{E}(\rho) \ketbra{x}{x}] = \bra{x} \mathcal{E}(\rho) \ket{x}$ and $\mathcal{P}_\textrm{ideal}(x) = \Tr[\rho_\textrm{ideal} \ketbra{x}{x}] = \bra{x}\rho_\textrm{ideal}\ket{x}$. If we take $\rho_\textrm{ideal} = \ketbra{0}{0}$, then $\mathcal{P}_\textrm{ideal}(0) = 1$ and $\mathcal{P}_\textrm{ideal}(1) = 0$. Therefore, we can write the TVD as
\begin{align}
    d_\textrm{TV}(\mathcal{P}, \mathcal{P}_\textrm{ideal}) 
        &= \frac{1}{2} \sum_{x \in \{0,1\}} |\mathcal{P}(x) - \mathcal{P}_\textrm{ideal}(x)| \\
        &= \frac{1}{2} \big[ |\mathcal{P}(0) - \mathcal{P}_\textrm{ideal}(0)| + |\mathcal{P}(1) - \mathcal{P}_\textrm{ideal}(1)| \big] \\
        &= \frac{1}{2} \big[ (1 - \mathcal{P}(0)) + \mathcal{P}(1) \big]  \\
        &= \mathcal{P}(1)  \\
        &= \bra{1} \mathcal{E}(\rho) \ket{1}.
\end{align}
In this case, the TVD is linear in $\rho$, and when $\mathcal{E}(\rho)$ is the sum over many randomizations under RC, we can write
\begin{align}
    d_\textrm{TV}(\mathcal{P}, \mathcal{P}_\textrm{ideal}) 
        &= \bra{1} \sum_i^N \mathcal{E}_i(\rho) \ket{1} \\
        &= \sum_i^N \bra{1} \mathcal{E}_i(\rho) \ket{1},
\end{align}
where the TVD is now simply a sum over $N$ independent noisy results, all of which are incorrect.

It may seem odd that the performance of RC depends on the linearity of an error metric to $\rho$; however, there is a physical interpretation for this phenomenon. Linear metrics, such as the fidelity or the TVD of deterministic algorithms, are only sensitive to the diagonal elements of the PTM error process, and are therefore insensitive to the off-diagonal terms produced by coherent errors; therefore, they will not benefit from the noise tailoring of these off-diagonal terms via RC. To understand this, we provide several examples below which also connect the current discussion with the discussion of error rates in the previous two sections. 

Consider the case in which $\rho_\textrm{ideal} = \ketbra{0}{0}$, but the actual qubit state has over-rotated by an angle $\theta$ due to coherent errors: $\ket{\psi} = \cos(\theta)\ket{0} + \sin(\theta)\ket{1}$. The fidelity of $\rho$ with $\rho_\textrm{ideal}$ is
\begin{equation}
    \mathcal{F} = |\braket{0}{\psi}|^2 = \cos^{2}(\theta) \approx (1 - \theta^2)^2,
\end{equation}
where we have approximated $\cos(\theta)$ in the small angle limit. Here, we can see that to first order the infidelity scales as $r(\mathcal{E}) \simeq \theta^2$. We can similarly compute the trace distance between $\rho$ and $\rho_\textrm{ideal}$,
\begin{align}
    D(\rho, \rho_\textrm{ideal}) &= \frac{1}{2} \Tr \big| \ketbra{\psi}{\psi} - \ketbra{0}{0} \big| \\
        & = \frac{1}{2} \Tr \Bigg| \begin{pmatrix} \cos^{2}(\theta) - 1 & \cos(\theta) \sin(\theta) \\  \cos(\theta) \sin(\theta)     & \sin^{2}(\theta) \end{pmatrix} \Bigg|  \\
        & = \sin^{2}(\theta) \approx \theta^2.
\end{align}
Similar to the fidelity, we can see that $r(\mathcal{E}) \simeq \theta^2$ for the trace distance when the target state is an eigenstate of the measurement basis. In this scenario, the trace distance between $\rho$ and $\rho_\textrm{ideal}$ is dual to the fidelity (i.e., overlap) of the two, but neither would benefit from RC in the many-randomization limit.

Next, consider the case in which the target state is in a superposition state, $\rho_\textrm{ideal} = \ketbra{+}{+}$, and the actual qubit state has over-rotated due to a small rotation about the y-axis,
\begin{align}
    \ket{\psi} & = R_y(\theta) \ket{+} \\
        & = \frac{1}{\sqrt{2}} \begin{pmatrix} \cos(\theta)  -\sin(\theta) \\  \sin(\theta) + \cos(\theta) \end{pmatrix}.
\end{align}
The fidelity of this state is
\begin{equation}
    \mathcal{F} = |\braket{+}{\psi}|^2 = \cos^{2}(\theta) \approx (1 - \theta^2)^2,
\end{equation}
where again we see that $\mathcal{F}$ is insensitive to the off-diagonal terms due to $R_y(\theta)$ and that the infidelity scales as $r(\mathcal{E}) \simeq \theta^2$. This is, however, not the case for the trace distance, where
\begin{align}
    D(\rho, \rho_\textrm{ideal}) &= \frac{1}{2} \Tr \big| \ketbra{\psi}{\psi} - \ketbra{+}{+} \big| \\
        & = \frac{1}{4} \Tr \Bigg| \begin{pmatrix} - 2\cos(\theta)\sin(\theta) & \cos^{2}(\theta) - \sin^{2}(\theta) - 1 \\ \cos^{2}(\theta) - \sin^{2}(\theta) - 1 & 2\cos(\theta)\sin(\theta) \end{pmatrix} \Bigg|  \\
        & = \cos(\theta)\sin(\theta) \approx \theta.
\end{align}
Therefore, we can see that when the target state is coherently spread out among the various basis states, the trace distance scales as $\sqrt{r(\mathcal{E})} \simeq \theta$, which is quadratically larger than the trace distance for deterministic algorithms. Consequently, the trace distance (and thus the TVD) can benefit greatly from the $\sqrt{r(\mathcal{E})} \longrightarrow r(\mathcal{E})$ reduction in the error under RC for non-deterministic algorithms.

The dependence of the error metric to the linearity of $\rho$ and the sensitivity to off-diagonal terms in the error process is demonstrated extensively in the data presented in this paper. For example, in Fig.~\ref{fig:Figure_S_QFT_improvement_exp} we show that the relative improvement under RC tends to $1$ for singular (peaked) output distributions. Furthermore, additional analyses of the single-qubit state tomography results provided in the supplement shows that the bare and RC TVDs are approximately equal when the target state is an eigenstate of the measurement basis (see Table \ref{Tab:Table_S_Tomography_TVD} and Fig.~\ref{fig:Figure_S_Tomography}), and that RC provides no general benefit when computing the fidelity of a noisy quantum state with the ideal quantum state (see Fig.~\ref{fig:Figure_S_Purity_Fidelity_vs_K}).

\section{TVD vs. Expectation Values in Quantum Algorithms}
\noindent
Most NISQ algorithms utilized measured expectation values for computing energies or determining wave-functions. In this work, we benchmarked the performance of RC using the TVD. However, the TVD provides an upper bound on the absolute error of the measured expectation values, which we outline below.

The expectation value of an operator $A$ acting on a quantum state $\rho$ is given as
\begin{equation}
    \expval{A}_{\rho} = \Tr[\rho A] = \sum_x \mathcal{P}(x) \bra{x} A \ket{x} = \sum_x \mathcal{P}(x) \expval{A}_{x},
\end{equation}
where we have written the $\rho$ in terms of its spectral decomposition, as described in the previous section, and $\expval{A}_{x}$ is the expectation value of $A$ in the $\ket{x}$ basis. We can write the absolute error in the expectation value as
\begin{equation}
    | \expval{A}_\rho - \expval{A}_{\rho, \textrm{true}} | = \Big| \sum_x (\mathcal{P}(x) - \mathcal{P}_\textrm{true}(x)) \expval{A}_x \Big|.
\end{equation}
It is then possible to bound the absolute error of this expectation value with a simple triangle inequality,
\begin{align}
    | \expval{A}_\rho - \expval{A}_{\rho, \textrm{true}} | 
        & \leq \sum_x \big| (\mathcal{P}(x) - \mathcal{P}_\textrm{true}(x)) \big| \big| \expval{A}_x \big| \\
        & \leq 2 d_\textrm{TV}(\mathcal{P}, \mathcal{P}_\textrm{true}) \norm{A},
\end{align}
where $\norm{A}$ is the operator norm of $A$, which is defined to be the maximal absolute value of all eigenvalues of $A$. Therefore, we can say that the absolute error in the expectation value of an arbitrary observable is upper-bounded by the TVD computed from the same probability distribution. While a reduction in the TVD under RC does not guarantee a reduction in the error of an expectation value measured in the same basis, since the error in the expectation value is only at most given by the TVD, it does tighten the bound within which one can accurately estimate the true expectation value. Since multiple expectation values can be estimated from any given probability distribution, the TVD effectively condenses information about the errors across all the relevant expectation values. In other words, if $d_\textrm{TV}(\mathcal{P}, \mathcal{P}_\textrm{true}) < \epsilon$, then the expectation value of all local observables will be accurate to within $\epsilon$. We therefore generally expect RC to improve the accuracy of expectation values in scenarios in which it improves the TVD.

\section{Single-qubit State Tomography}

\begin{figure}
\centerline{\includegraphics[width=0.95\textwidth]{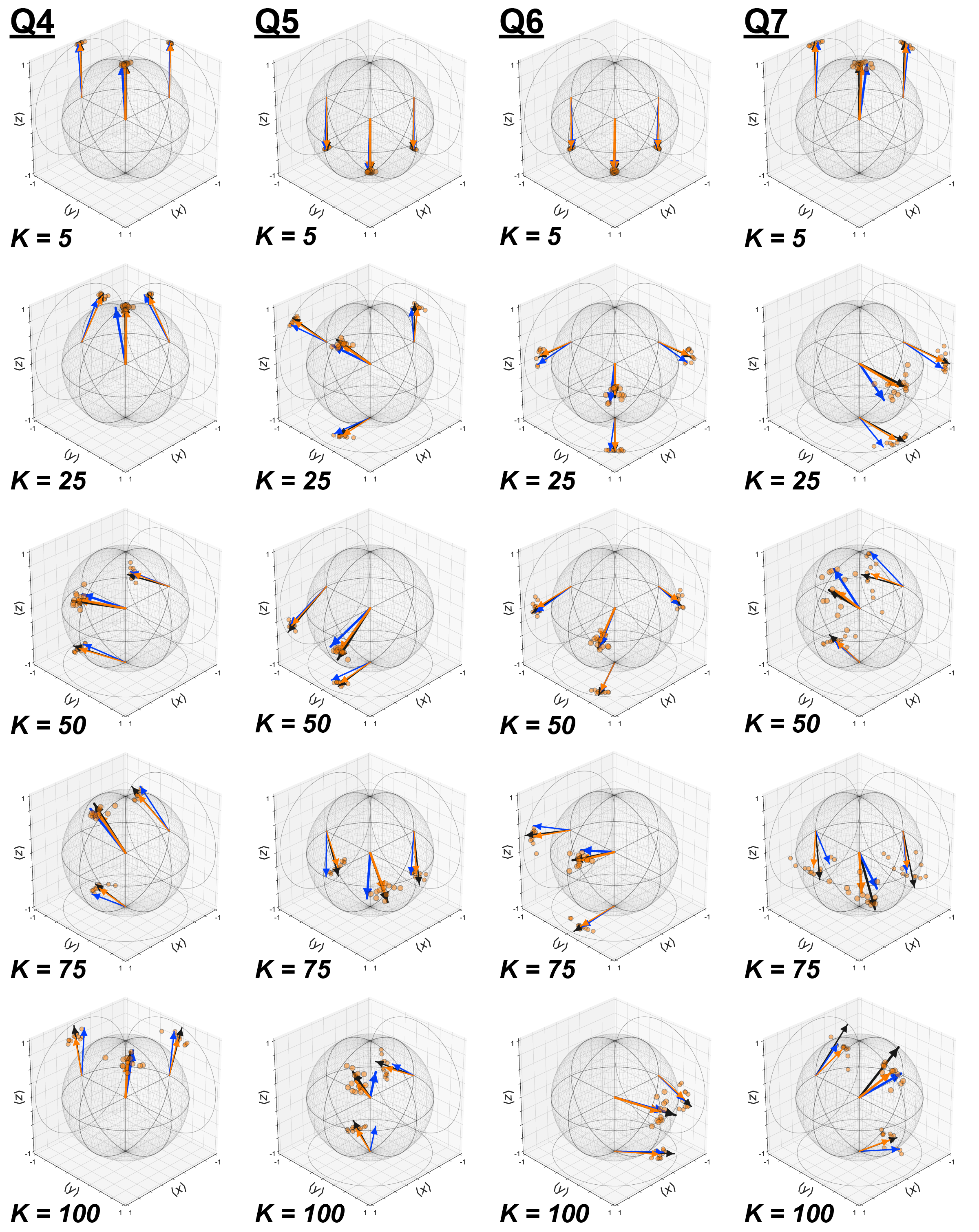}}
\caption{
    \textbf{Single-qubit state tomography results.} $K$ is the circuit depth (number of cycles of easy/hard gates). The black vector is the ideal result; the blue vector is the bare result; the orange points are the individual RC results; and the orange vector is the combined RC result.}
    \label{fig:Figure_S_Tomography}
\end{figure}

\begin{table}
\centering
\resizebox{0.8\textwidth}{!}{
    \begin{tabular}{|l||c|c||c|c||c|c||c|c||c|c|}
    	\hline
    	\multirow{2}{*}{\textbf{Qubit (Depth)}} & \multicolumn{2}{|c||}{\boldmath$\textbf{TVD}_x$} & \multicolumn{2}{|c||}{\boldmath$\textbf{TVD}_y$} & \multicolumn{2}{|c||}{\boldmath$\textbf{TVD}_z$} & \multicolumn{2}{|c||}{\textbf{Purity}} & \multicolumn{2}{|c|}{\textbf{Fidelity}} \\
    	\cline{2-11}
    	& \textbf{Bare} & \textbf{RC} & \textbf{Bare} & \textbf{RC} & \textbf{Bare} & \textbf{RC} & \textbf{Bare} & \textbf{RC} & \textbf{Bare} & \textbf{RC} \\
    	\hline
    	
    	\hline
        \textbf{Q4 (\boldmath$K = 5$)} & 0.031(8) & 0.006(2) & 0.009(8) & 0.002(2) & 0.009(2) & 0.0089(4) & 0.985 & 0.982 & 0.983 & 0.982 \\
        \hline
        \textbf{Q4 (\boldmath$K = 25$)} & 0.074(8) & 0.014(2) & 0.037(7) & 0.020(2) & 0.029(6) & 0.024(2) & 0.935 & 0.932 & 0.922 & 0.932 \\
        \hline
        \textbf{Q4 (\boldmath$K = 50$)} & 0.077(8) & 0.004(2) & 0.037(7) & 0.062(1) & 0.042(8) & 0.035(2) & 0.891 & 0.870 & 0.879 & 0.869 \\
        \hline
        \textbf{Q4 (\boldmath$K = 75$)} & 0.130(8) & 0.024(2) & 0.095(6) & 0.072(2) & 0.038(7) & 0.049(2) & 0.887 & 0.821 & 0.838 & 0.820 \\
        \hline
        \textbf{Q4 (\boldmath$K = 100$)} & 0.111(8) & 0.010(2) & 0.081(8) & 0.038(2) & 0.060(4) & 0.107(2) & 0.845 & 0.773 & 0.812 & 0.772 \\
        \hline
        
        \hline
        \textbf{Q5 (\boldmath$K = 5$)} & 0.029(8) & 0.008(2) & 0.011(8) & 0.006(2) & 0.020(2) & 0.021(1) & 0.963 & 0.958 & 0.961 & 0.957 \\
        \hline
        \textbf{Q5 (\boldmath$K = 25$)} & 0.014(6) & 0.013(2) & 0.034(8) & 0.033(2) & 0.059(7) & 0.030(2) & 0.945 & 0.941 & 0.937 & 0.939 \\
        \hline
        \textbf{Q5 (\boldmath$K = 50$)}) & 0.025(5) & 0.042(2) & 0.104(8) & 0.023(2) & 0.060(8) & 0.046(2) & 0.892 & 0.876 & 0.868 & 0.875 \\
        \hline
        \textbf{Q5 (\boldmath$K = 75$)} & 0.138(8) & 0.003(2) & 0.097(8) & 0.035(2) & 0.054(4) & 0.090(1) & 0.852 & 0.818 & 0.800 & 0.815 \\
        \hline
        \textbf{Q5 (\boldmath$K = 100$)} & 0.067(6) & 0.073(2) & 0.195(7) & 0.106(2) & 0.050(8) & 0.015(2) & 0.782 & 0.748 & 0.716 & 0.746 \\
        \hline
        
        \hline
        \textbf{Q6 (\boldmath$K = 5$)} & 0.009(8) & 0.009(2) & 0.010(8) & 0.003(2) & 0.023(2) & 0.023(1) & 0.955 & 0.955 & 0.955 & 0.955 \\
        \hline
        \textbf{Q6 (\boldmath$K = 25$)} & 0.010(6) & 0.024(2) & 0.041(6) & 0.020(2) & 0.062(8) & 0.009(2) & 0.967 & 0.939 & 0.957 & 0.939 \\
        \hline
        \textbf{Q6 (\boldmath$K = 50$)} & 0.038(5) & 0.046(2) & 0.020(7) & 0.027(2) & 0.009(8) & 0.029(2) & 0.918 & 0.887 & 0.918 & 0.886 \\
        \hline
        \textbf{Q6 (\boldmath$K = 75$)} & 0.088(5) & 0.087(1) & 0.003(8) & 0.015(2) & 0.052(8) & 0.045(2) & 0.874 & 0.807 & 0.861 & 0.806 \\
        \hline
        \textbf{Q6 (\boldmath$K = 100$)} & 0.045(7) & 0.091(2) & 0.077(7) & 0.077(2) & 0.020(8) & 0.039(2) & 0.821 & 0.752 & 0.820 & 0.751 \\
        \hline
        
        \hline
        \textbf{Q7 (\boldmath$K = 5$)} & 0.037(8) & 0.006(2) & 0.044(8) & 0.018(2) & 0.012(2) & 0.017(1) & 0.983 & 0.968 & 0.977 & 0.967 \\
        \hline
        \textbf{Q7 (\boldmath$K = 25$)} & 0.170(8) & 0.029(2) & 0.069(4) & 0.060(1) & 0.073(8) & 0.008(2) & 0.938 & 0.881 & 0.862 & 0.879 \\
        \hline
        \textbf{Q7 (\boldmath$K = 50$)} & 0.041(8) & 0.031(2) & 0.105(5) & 0.114(2) & 0.251(8) & 0.052(2) & 0.844 & 0.754 & 0.705 & 0.751 \\
        \hline
        \textbf{Q7 (\boldmath$K = 75$)} & 0.113(8) & 0.068(2) & 0.090(8) & 0.084(2) & 0.107(5) & 0.149(2) & 0.808 & 0.672 & 0.762 & 0.658 \\
        \hline
        \textbf{Q7 (\boldmath$K = 100$)} & 0.112(7) & 0.065(2) & 0.136(7) & 0.017(2) & 0.139(8) & 0.210(2) & 0.761 & 0.657 & 0.688 & 0.619 \\
        \hline
    \end{tabular}}
\caption{
\textbf{Total variation distance, purity, and fidelity of the single-qubit state tomography results.} The TVD is calculated with respect to each measurement basis for all qubits at each circuit depth. The state fidelity and purity are calculated for the bare results and unioned RC results.}
\label{Tab:Table_S_Tomography_TVD}
\end{table}

\begin{figure}
\centerline{\includegraphics[width=0.8\textwidth]{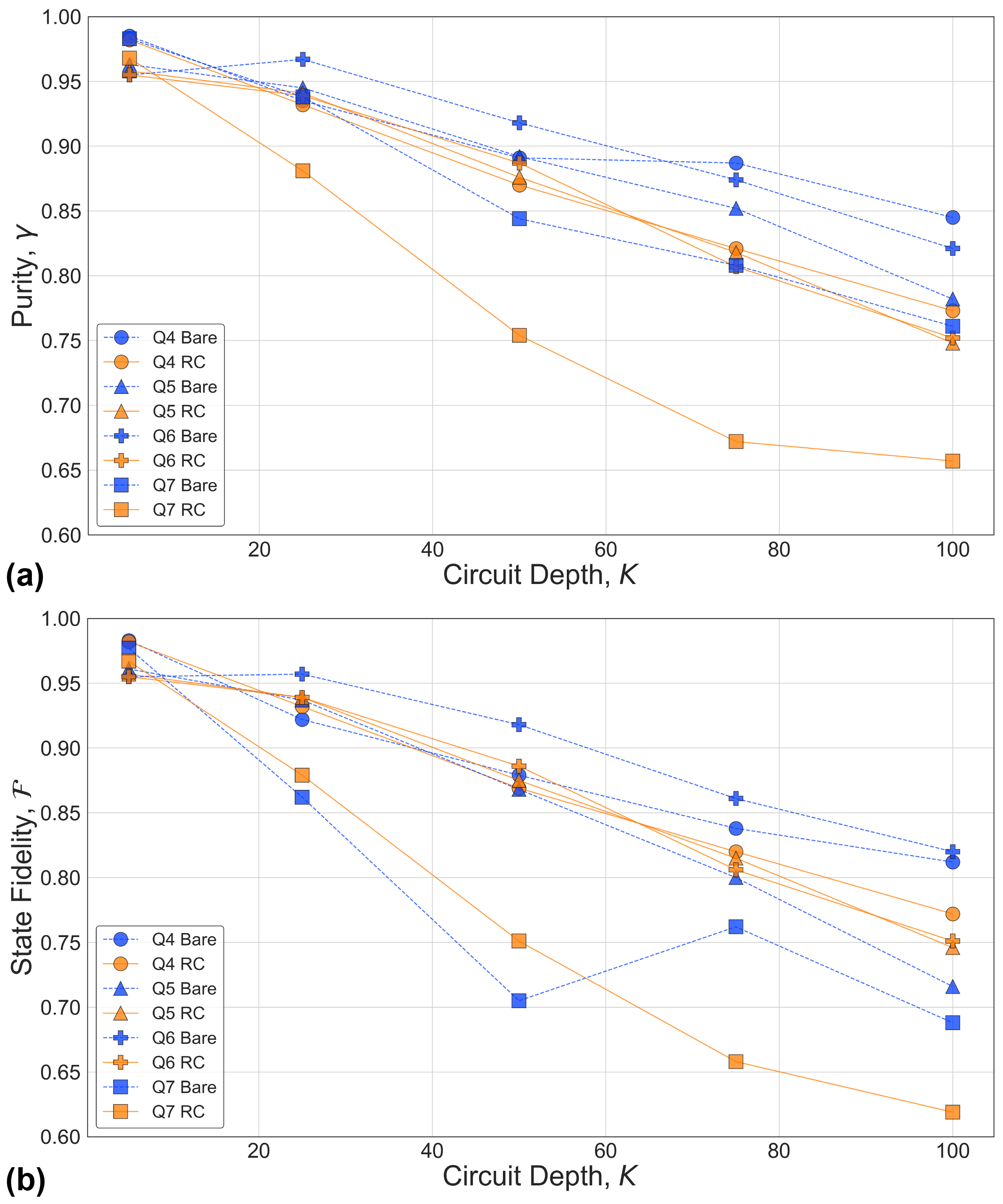}}
\caption{
    \textbf{Purity and State Fidelity vs.~Circuit Depth for the single-qubit state tomography results.} The purity and fidelity of the bare results can differ significantly due to angle errors in the bare states resulting from coherent errors. In contrast, the purity and fidelity of the RC results are approximately equal in magnitude at each circuit depth, demonstrating that the most dominant error mechanism under RC is stochastic noise.}
    \label{fig:Figure_S_Purity_Fidelity_vs_K}
\end{figure}

\noindent
Single-qubit dynamics provide an intuitive picture for understanding how noise is tailored under RC: consider the ideal final state of a qubit after performing a sequence of gates. The ensemble measurements of this final state will yield a probability distribution of 0s and 1s, which depends on the location of the final state on the Bloch sphere and the measurement basis. In a circuit dominated by coherent errors, the actual final state of a ``bare'' circuit will have some angle error relative to the ideal state due to the interference of coherent errors during the sequence of gates. Under RC, the impact of coherent errors is different for each randomization, since each circuit represents a slightly different trajectory to the same (ideal) final state. Thus, the combined distribution of all the randomizations averages out the impact of coherent errors, resulting in a stochastic Pauli channel (Eq.~\ref{sec_cb_cer}.\ref{pauli_channel}).

To verify this intuitive picture of noise tailoring via RC, we performed state tomography on all four qubits independently, with and without RC. Random circuits were generated by randomly sampling $K=100$ interleaved cycles of ``easy'' and ``hard'' single-qubit gates, as defined by the following gate sets: the Clifford set, $C_{easy} = \{ \mathbf{C_1} \}$, and common non-Clifford gates, $G_{hard} = \{ X45, \; Y45, \; T=Z45 \}$, respectively. Thus, the total number of single-qubit gates for each qubit was $2K + 1 = 201$. State tomography results were constructed by performing ensemble measurements in the $X$, $Y$, and $Z$ bases at five different points during each random circuit, as defined by the circuit depth: $K=5$, $K=25$, $K=50$, $K=75$, and $K=100$. These results are plotted in Fig.~\ref{fig:Figure_S_Tomography}; 2d projections of the results are also included for visual clarity. At each circuit depth, measurements were performed on the bare circuit (blue vector) and on 12 different randomizations (orange points). The ideal result is plotted as a black vector, and the combined RC result is represented by the orange vector. 6,000 shots were taken in each measurement basis for the bare circuits, and 500 shots were taken for each randomization. Focusing on Q5 as an example, at the beginning of the circuit ($K=5$) all three vectors are more or less co-aligned at the south pole. However, at later circuit depths, coherent errors manifest as a separation between the ideal and bare vectors. At all circuit depths, the RC vector is approximately co-linear with the ideal state.

Since each tomography result corresponds to measuring same final state in $X$, $Y$, and $Z$, the TVD can be calculated with respect to each basis. These results are summarized in Table \ref{Tab:Table_S_Tomography_TVD}. When the target state is approximately aligned with the measurement basis, as is the case for all four qubits at $K = 5$ with respect to $Z$, the bare and RC TVD performance is approximately equal. This is a reflection of the fact that RC is not expected to provide algorithmic improvement when the target state is an eigenstate of the measurement basis.

The consequence of tailoring coherent errors into stochastic Pauli errors is that the tailored noise becomes a decoherence channel, resulting in a reduction of the purity of the Bloch vector. The purity and state fidelity of each result are included in Table \ref{Tab:Table_S_Tomography_TVD}, and these results are summarized in Fig.~\ref{fig:Figure_S_Purity_Fidelity_vs_K}, in which both are plotted as a function of circuit depth for all four qubits. As seen in this plot, the purities of the RC vectors decrease more rapidly than their bare counterparts. However, the state fidelities of the RC results are approximately equal in magnitude to their respective purities, underscoring that the most dominant error mechanism under RC is stochastic noise. Under pure depolarizing noise, the RC vector would decrease linearly with circuit depth; the deviation from perfect linearity in our case shows that the tailored noise under RC is due to Pauli errors with different contributions. While both the purity and state fidelity decrease monotonically for the RC results, the same is not always true for the bare results due to the nature of coherent errors and how they interfere over the course of a circuit. As an example, we see the bare fidelity of Q7 increases from $K=50$ to $K=75$, even though the corresponding purity continues to decrease.

\section{Quantum Fourier Transform}

\begin{figure}
\centerline{\includegraphics[width=\textwidth]{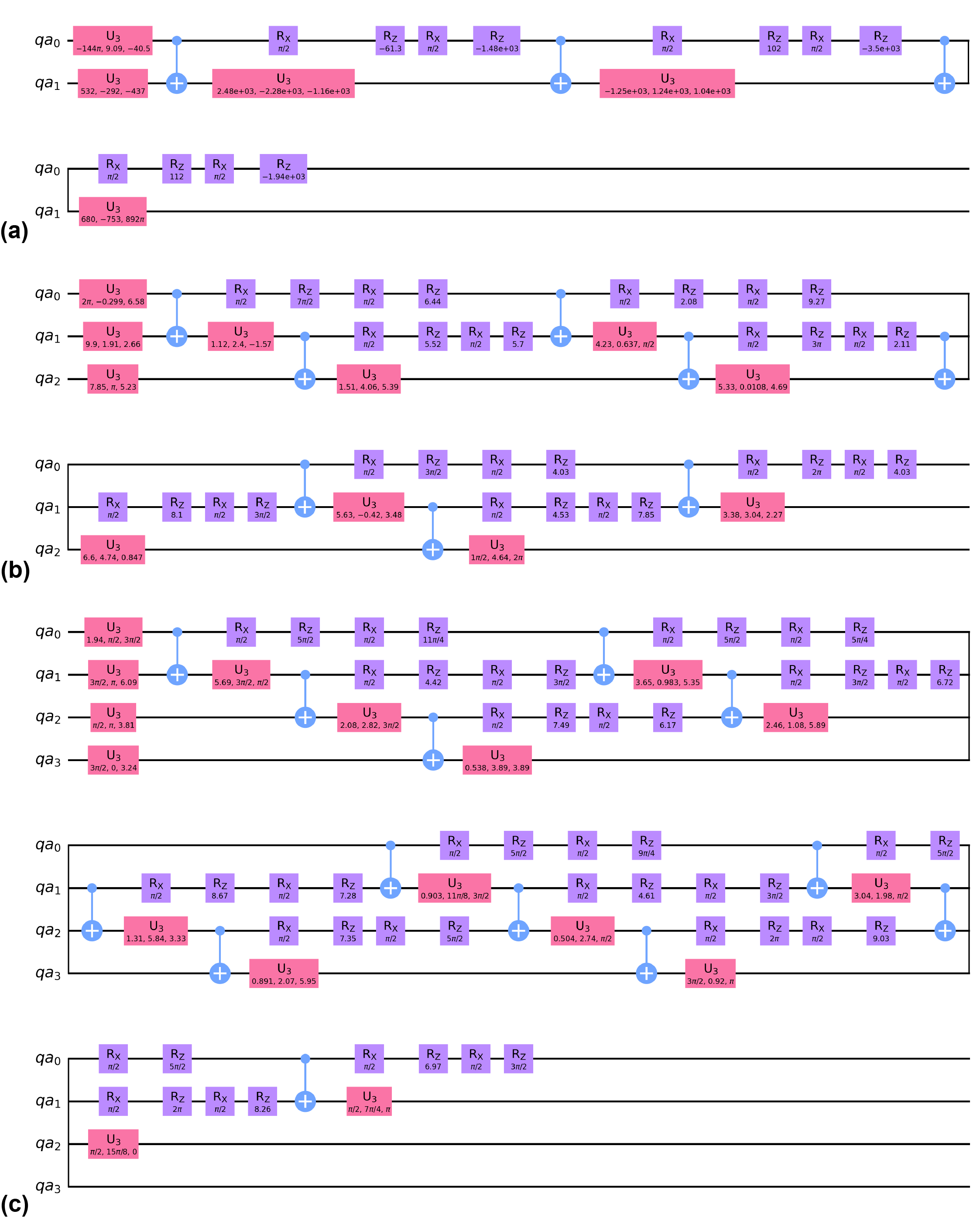}}
\caption{
    \normalfont{Optimal decompositions for the (\textbf{a}) two-, (\textbf{b}) three-, (\textbf{c}) and four-qubit quantum Fourier transform algorithms used in this work.}}
    \label{fig:Figure_S_QFT_circuit_diagrams}
\end{figure}

\begin{figure}
\centerline{\includegraphics[width=0.68\textwidth]{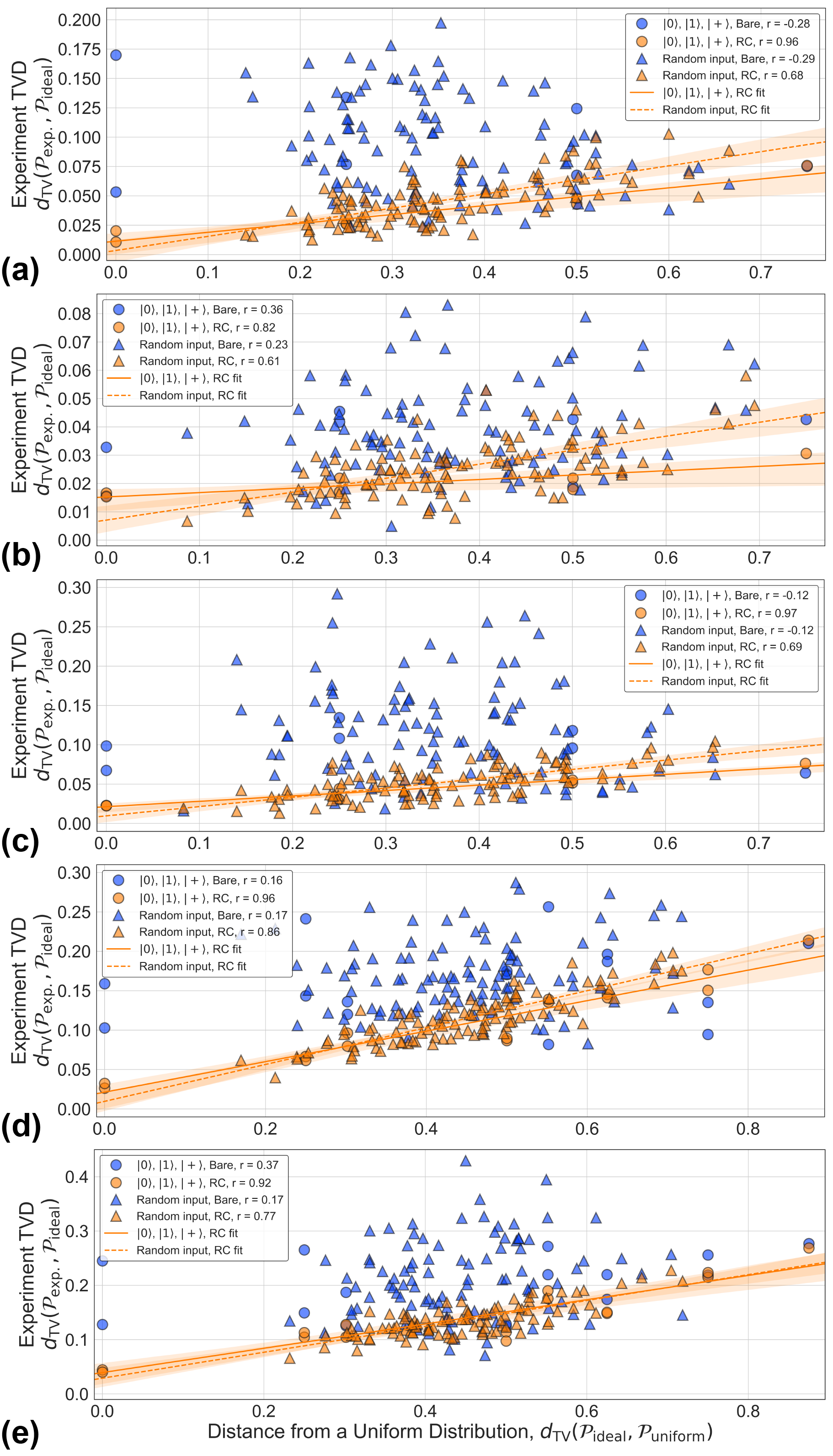}}
\caption{
    \normalfont{Two- and three-qubit quantum Fourier transform results for (\textbf{a}) Q4 \& Q5, (\textbf{b}) Q5 \& Q6, (\textbf{c}) Q6 \& Q7, (\textbf{d}) Q4, Q5, \& Q6, and (\textbf{e}) Q5, Q6, \& Q7.} Pearson $r$ values listed in the legend indicate the linear correlation of each data set. Linear fits are plotted for the RC data, with transparent bands indicating the 95\% confidence intervals.}
    \label{fig:Figure_S_QFT_2_3_qubits}
\end{figure}

\begin{figure}
\centerline{\includegraphics[width=0.9\textwidth]{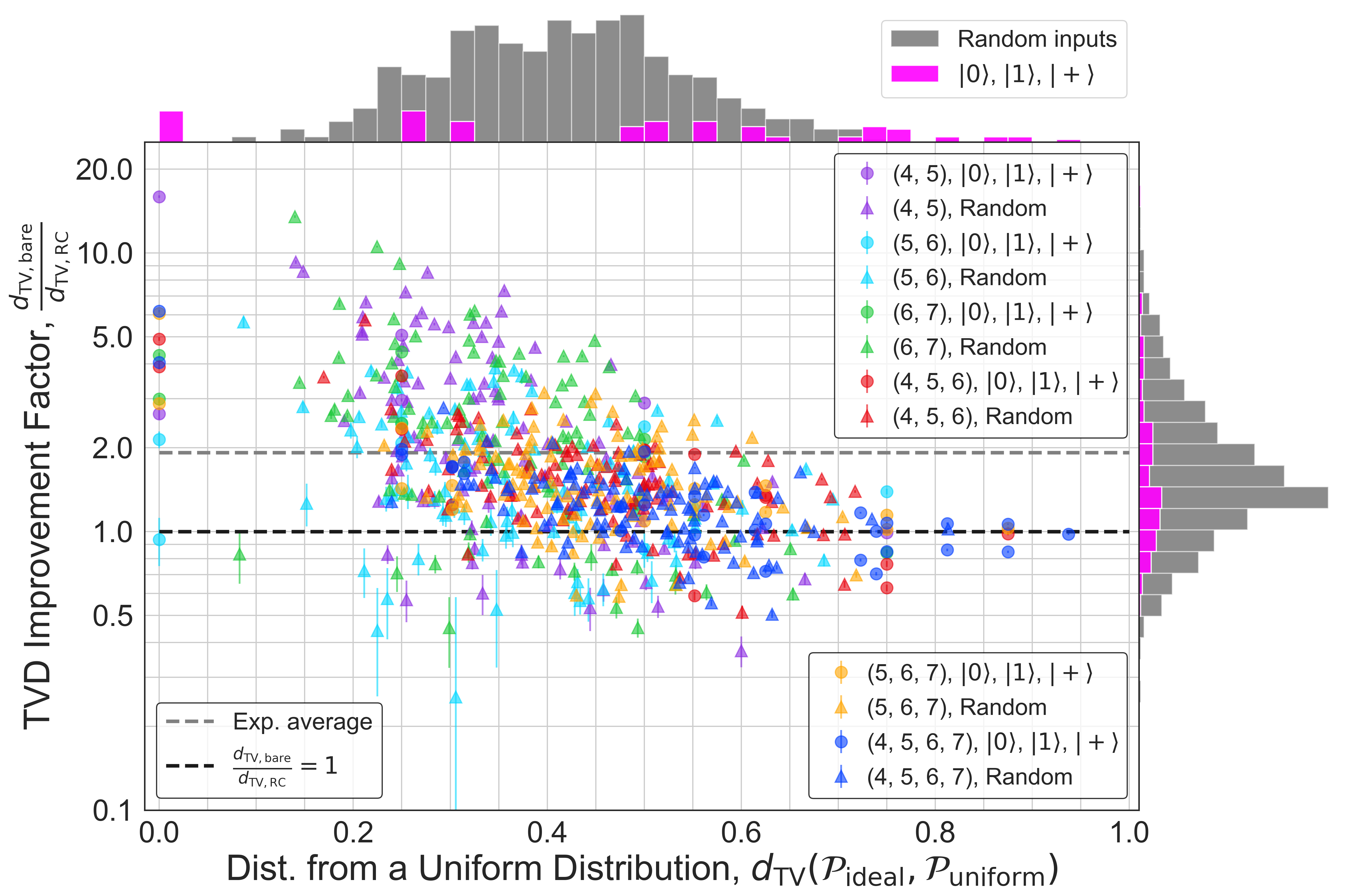}}
\caption{
    \textbf{Experimental results of the TVD improvement under RC for the quantum Fourier transform applied to two, three, and four qubits, as a function of the distribution uniformity of the ideal results.} The qubit subsets are specified in the legend. Histograms of the Pauli input (pink) and random input (grey) results are included on both axes to show how the results are distributed. RC performs better for results in which $d_\textrm{TV,bare} /d_\textrm{TV,RC} > 1$, but performs worse for results in which $d_\textrm{TV,bare} /d_\textrm{TV,RC} < 1$. For all random input results, the average improvement is $d_\textrm{TV,bare} /d_\textrm{TV,RC} = 1.92$.}
    \label{fig:Figure_S_QFT_improvement_exp}
\end{figure}

\noindent
In this work, we utilized a state-of-the-art synthesis algorithm \cite{davis2019heuristics} that numerically approximates circuit unitaries in order to reduce the CNOT count for a given hardware topology, which generated two-, three-, and four-qubit QFT circuits consisting of only $K=3$, $K=8$, and $K=13$ CNOTs, respectively, for our linear connectivity (see Fig.~\ref{fig:Figure_S_QFT_circuit_diagrams}). Experimental TVD results for the two- and three-qubit QFT algorithms can be seen in Fig.~\ref{fig:Figure_S_QFT_2_3_qubits}. For each qubit subset, single-qubit product states involving permutations of $\ket{0}$, $\ket{1}$, and $\ket{+}$ constitute the Pauli input states, and ``random'' input states were generated by applying random $SU(2)$ unitaries to each qubit independently before applying the QFT algorithm. $N = 50$ randomizations were generated for each bare circuit, and 100 random inputs were generated for each data set. Each bare circuit was measured 10,000 times, and each randomization was measured 200 times. A summary of the TVD improvement under RC for all two-, three-, and four-qubit results can be seen in Fig.~\ref{fig:Figure_S_QFT_improvement_exp}, plotted as the ratio of the bare to RC TVDs, $d_\textrm{TV,bare} / d_\textrm{TV,RC}$, as a function of the distribution uniformity of the ideal results. The histogram of the TVD improvement under RC (y-axis) for the random input results is included in Fig.~3e of the main text.

Simulated QFT results were generated using two models: (1) a Pauli model, using the Pauli error rates directly extracted from the CER results in Fig.~\ref{fig:Figure_S_KNR}(b); (2) a complete model of our system that includes coherent errors, which produces simulated CER results that are approximately equal to the error rates in Fig.~\ref{fig:Figure_S_KNR}(b). The complete model of our four-qubit system is generated by finding a desired CPTP map for a given unitarity by minimizing simulated CER results with respect to the experimental results; we only considered single-body errors in the construction of this model, as these are the most dominant error syndromes in our system (more details can be found in the next section). Our simulator takes into account realistic values for measurement errors (see Table \ref{tab:sqp}) and implements coherent and stochastic errors in each cycle of a circuit. Stochastic noise acts on each qubit per cycle with a finite probability, and coherent errors are implemented by adding an over-rotation $\theta$ to each qubit (Eq.~\ref{sec_coherent_error}.\ref{coherent_error}), where $\theta$ is set according to the process infidelity due to coherent errors. A similar model is used for simulating coherent errors on two-qubit gates. The histograms of the TVD improvement for random input states under RC for the simulated results (using the complete model) in Figs.~\ref{fig:Figure_S_QFT_improvement_sim}, \ref{fig:Figure_S_QFT_improvement_sim_scaled_1Q}, and \ref{fig:Figure_S_QFT_improvement_sim_scaled_both} are included in Fig.~3e of the main text.

\section{Simulation Model}

\begin{figure}
\centerline{\includegraphics[width=0.9\textwidth]{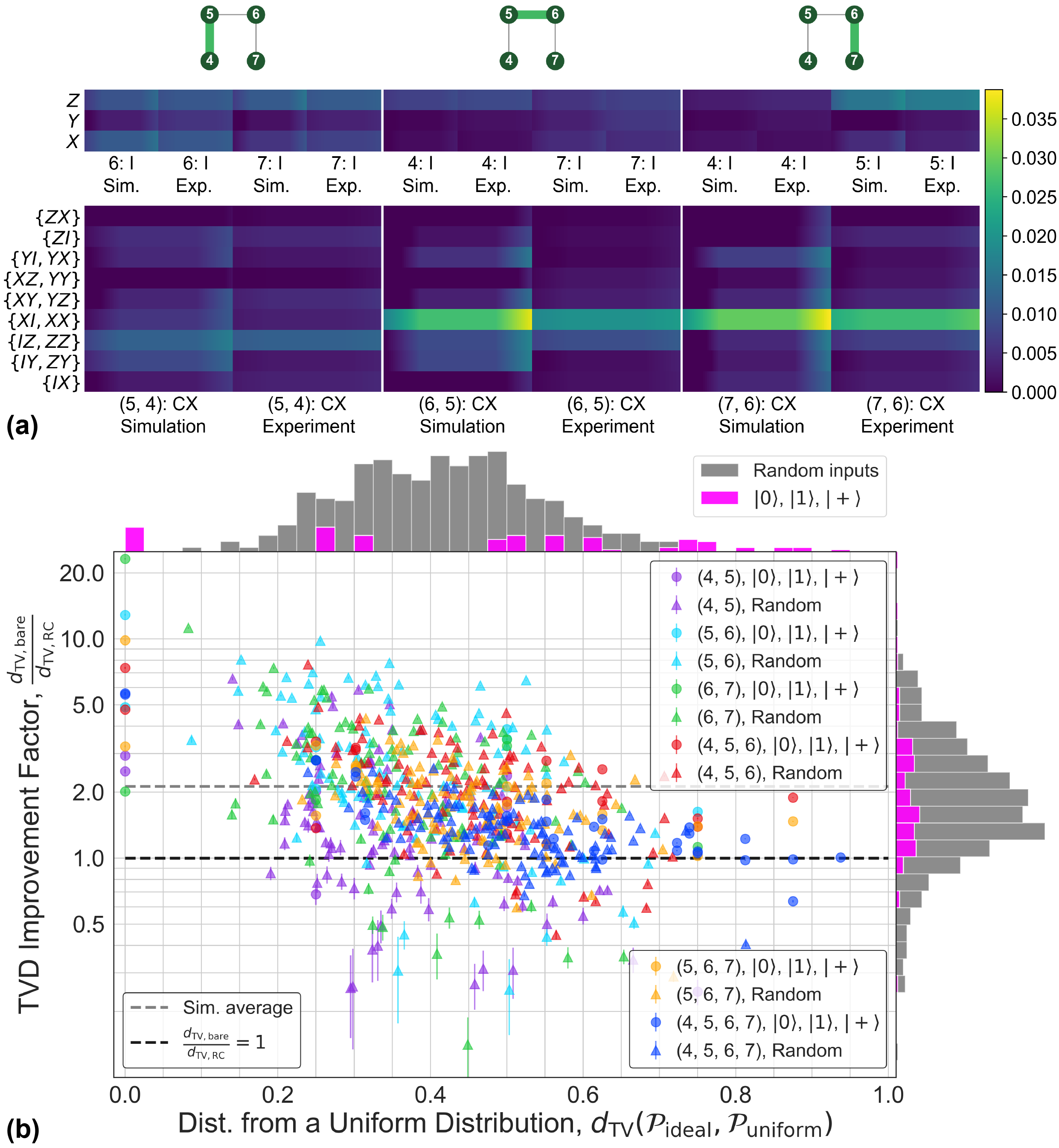}}
\caption{
    \textbf{Simulated QFT results using a complete model with equivalent error rates.}
    \textbf{(a)} \normalfont{Experimental vs.~simulated cycle error reconstruction results based on a model of our system with equivalent single-body error rates in which the fraction of the total error rate due to coherent errors was set to 0.7 (0.9) for single-qubit (two-qubit) gates.}
    \textbf{(b)} \normalfont{Simulated results for data in Fig.~\ref{fig:Figure_S_QFT_improvement_exp} using the model presented in \textbf{a}. The average improvement is $d_\textrm{TV,bare} /d_\textrm{TV,RC} = 2.13$ for all random input results, showing good agreement with the experimental results in Fig.~\ref{fig:Figure_S_QFT_improvement_exp}.} }
    \label{fig:Figure_S_QFT_improvement_sim}
\end{figure}

\begin{figure}
\centerline{\includegraphics[width=0.9\textwidth]{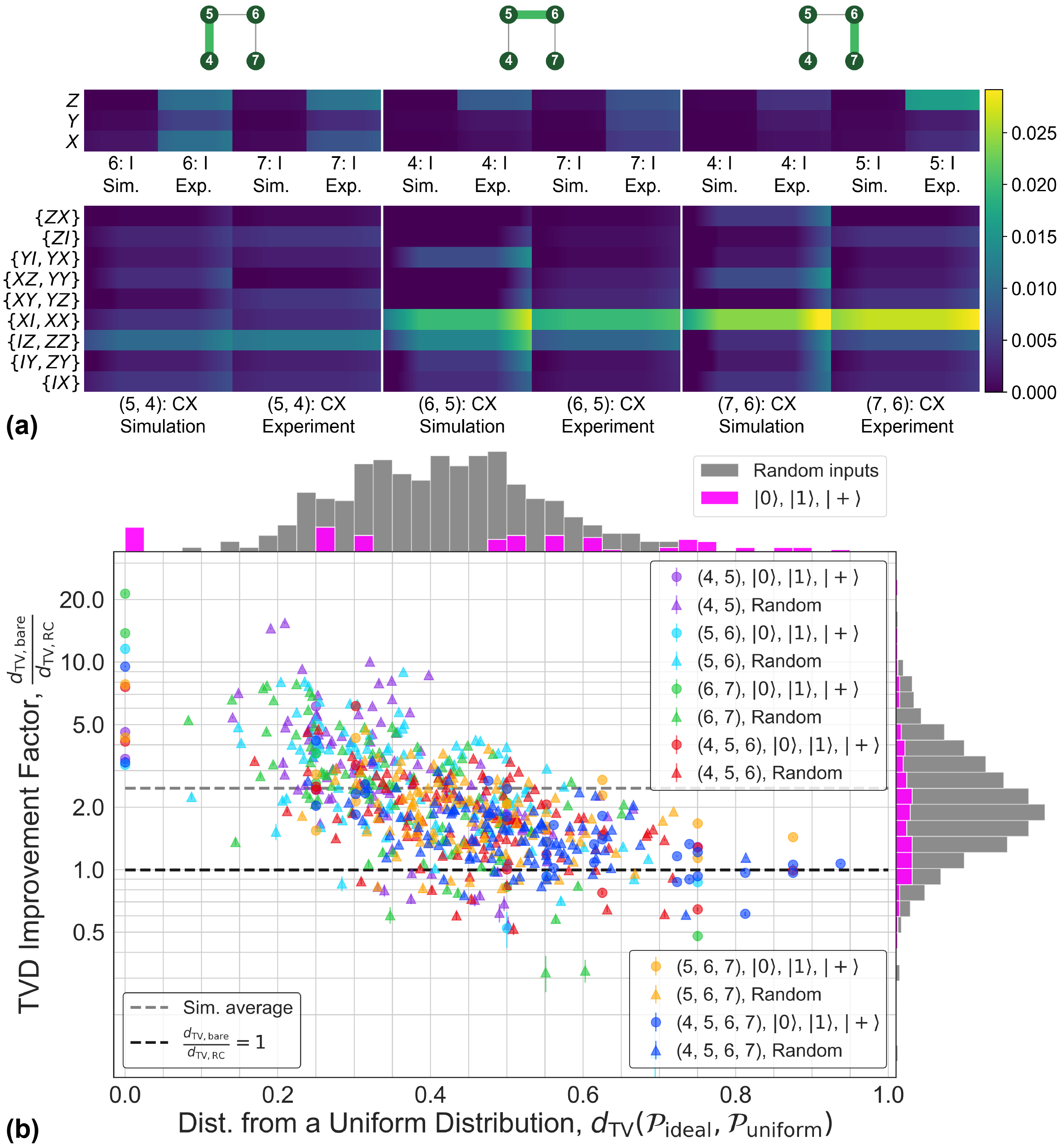}}
\caption{
    \textbf{Simulated QFT results using a complete model with improved single-qubit error rates.}
    \textbf{(a)} \normalfont{Experimental vs.~simulated cycle error reconstruction results, with the single-qubit error rates reduced by a factor of 10 compared to the  model presented in Fig.~\ref{fig:Figure_S_QFT_improvement_sim}.}
    \textbf{(b)} \normalfont{Simulated results for data in Fig.~\ref{fig:Figure_S_QFT_improvement_exp} using the model presented in \textbf{a}. The average improvement is $d_\textrm{TV,bare} /d_\textrm{TV,RC} = 2.47$ for all random input results.} }
    \label{fig:Figure_S_QFT_improvement_sim_scaled_1Q}
\end{figure}

\begin{figure}
\centerline{\includegraphics[width=0.9\textwidth]{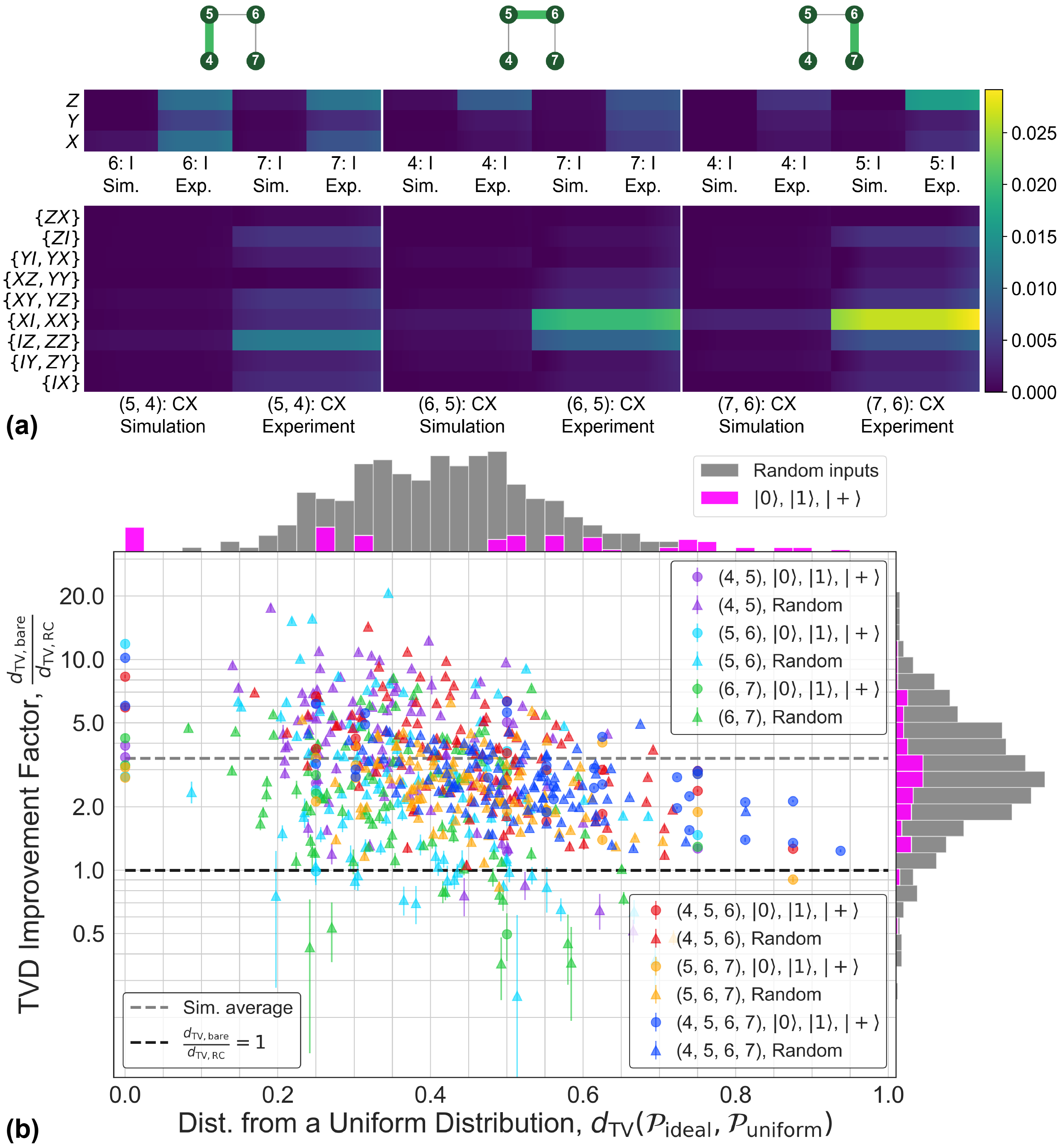}}
\caption{
    \textbf{Simulated QFT results using a complete model with improved single- and two-qubit error rates.}
    \textbf{(a)} \normalfont{Experimental vs.~simulated cycle error reconstruction results, with both the single- and two-qubit error rates reduced by a factor of 10 compared to the model presented in Fig.~\ref{fig:Figure_S_QFT_improvement_sim}.}
    \textbf{(b)} \normalfont{Simulated results for data in Fig.~\ref{fig:Figure_S_QFT_improvement_exp} using the model presented in \textbf{a}. The average improvement is $d_\textrm{TV,bare} /d_\textrm{TV,RC} = 3.39$ for all random input results.} }
    \label{fig:Figure_S_QFT_improvement_sim_scaled_both}
\end{figure}

\noindent
The circuits we are interested in simulating, as with those performed experimentally, have been structured to alternate between cycles containing CNOTs (hard cycles) and cycles of single qubit gates (easy cycles). Simulation comprises of two steps: first, we define a noise channel for every unique hard cycle in the collection of circuits to be simulated. Next, given a circuit, we propagate the initial state by simulating each easy cycle ideally, and each hard cycle ideally but followed by its corresponding noise channel. We do not add any noise to the easy cycles not because we are assuming they are perfect in practice, but rather because our noisy channels will be constructed using CER data, which measures the composition of the errors due to a cycle of random Pauli gates along with a particular hard cycle (see Eq.~\ref{sec_cb_cer}.\ref{cb_process_infidelity_CT}). We do, however, assume no correlations between the errors of each gate in a given cycle: we take the total noise channel of a cycle as the tensor product of noise channels of each gate or idling qubit. This assumption allows for noise induced by classical crosstalk (since our noise model is entirely cycle-dependent, we are explicitly allowing noise on one gate to be conditional on the existence of another gate in the cycle), but not noise induced by quantum crosstalk. Our justification is that our CER data shows no such significant correlations (see Fig.~\ref{fig:Figure_S_KNR}), so it does not appear necessary to include two-body terms in our model as the most dominant errors are single-body. Therefore, limiting ourselves to only single-body errors was sufficient for producing an approximate error model for our system, and simplified the process of finding the model. In general, measuring all $n$-body errors is manifestly not scalable as the number of such errors grows exponentially with $n$. Moreover, the probabilities of two-body errors are the sum of the probabilities of all errors that act non-trivially on the corresponding two bodies, irrespective of their action on other bodies. Therefore, the fact that two-body errors are negligible shows that three or more body errors are also negligible. In what follows, we describe the procedure to generate a noise channel for a specific gate or idling qubit in a given cycle.

First, given a probability simplex $q \in \mathbb{R}^N$ and a real vector $h\in\mathbb{R}^{N-1}$ we can define a quantum channel
\begin{equation}
    S(q,h) = \mathcal{U}_h\mathcal{K}_q,
\end{equation}
where $\mathcal{U}_h$ is the unitary superoperator corresponding to the unitary matrix
\begin{equation}
    U_h = \operatorname{exp}\left(-i \sum_{i=1}^{N-1} h_i P_i\right),
\end{equation}
and where $\mathcal{K}_q$ is the Pauli Kraus channel
\begin{equation}
    \mathcal{K}_q(\rho) = \sum_{i=0}^N \sqrt{q_i} P_i\rho P_i^\dagger.
\end{equation}
Here, $P_0,...,P_{N-1}$ with $N=4^n$ is some enumeration of the $n$-qubit Paulis such that $P_0=I$. $S(q,h)$ defines a somewhat arbitrary but large class of CPTP channels which is sufficient but not necessary to suit our needs: we require a parameterized class of CPTP channels over which to perform a numerical search.

Next, given any pair $(q,h)$ we can compute the Pauli transfer matrix (PTM) of $S(q,h)$, whose diagonal vector we denote as $d(q,h)$. We can likewise compute the unitarity of the channel $S(q,h)$ by taking the 2-norm of the lower PTM block \cite{wallman2015estimating}, denoting it by $u(q,h)$.

Finally, suppose that for some subset of qubits we experimentally measure the PTM diagonal to have a value of $f\in \mathbb{R}^N$. This is naturally done with cycle error reconstruction (see, for example the ``(5,4): CX'' block of the first cycle in Fig.~\ref{fig:Figure_S_KNR}), where the error Kraus probabilities are reported for the qubits (4, 5), from which the PTM diagonal for these qubits can be constructed via the inverse Walsh-Hadamard transform \cite{flammia2019efficient}. We wish to define a channel for simulation whose PTM diagonal matches $f$ up to a user-defined scaling, and such that the channel has a user-defined unitarity. Therefore, we perform a numerical optimization to find $(q,h)$ such that
\begin{align}
    d(q,h) &= 1 - s_0 (1 - f) \\
    u(q,h) &= 1 - (1 - s_1) (1 - \overline{f}^2)
\end{align}
where $s_0\in[0,1]$ defines the factor with which to decrease the process infidelity (recall the process infidelity corresponding to $f$ is $1-\overline{f}$ where $\overline{f}$ is the mean value of $f$), and where $s_1\in[0,1]$ defines the unitarity fraction, where a value of 1 results in $S(q,h)$ being unitary, and $0$ results in $S(q,h)$ being as stochastic as possible given the constraints. This minimization is performed with SciPy's BGFS solver for all non-overlapping subsets of qubits of the device, which is valid because the experimental data show no significant correlated error between gate-bodies. Minima are consistently found to within numerical precision, though minimum values are not unique.

The tensor product of the resulting channels, one for each gate or idling qubit in the cycle, defines the noisy channel for the given hard cycle. We allow a different value of $s$ for single- and two-qubit subsets. Guided by our CB results, RB, and unitary RB measurements, $s_1$ was set to 0.7 (0.9) for single-qubit (two-qubit) gates for all simulated results. $s_0 = 1.0$ for the results in Fig.~\ref{fig:Figure_S_QFT_improvement_sim} to simulate error rates that are equivalent to experimental values. However, $s_0 = 0.1 (1.0)$ for single-qubit (two-qubit) gates for the results in Fig.~\ref{fig:Figure_S_QFT_improvement_sim_scaled_1Q}, and $s_0 = 0.1$ for both single- and two-qubit gates for the results in Fig.~\ref{fig:Figure_S_QFT_improvement_sim_scaled_both}.

\section{Random Circuits of Variable Depth}

\begin{figure}
\centerline{\includegraphics[width=\textwidth]{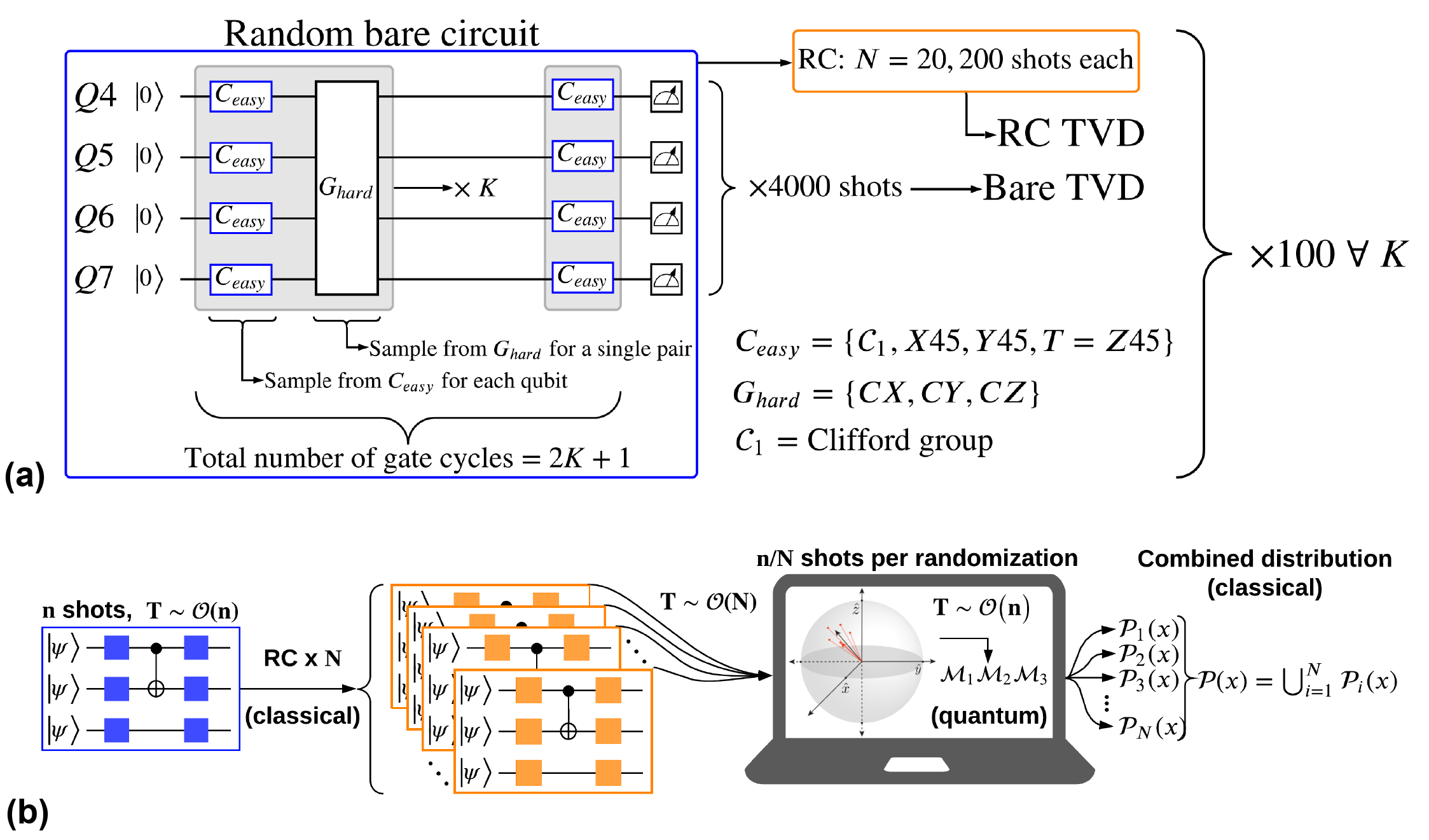}}
\caption{
    \textbf{Random circuit sampling.
    \textbf{(a)} \normalfont{The process by which $K$ cycles of interleaved easy/hard gates are randomly sampled from a universal gate set.}
    \textbf{(b)} \normalfont{Measurements using randomized compiling. Use the RC randomization protocol to generate $N$ randomizations of a bare circuit. Measure each randomization $n / N$ times, then combine all the results to obtain an equivalent statistical distribution for a circuit measured $n$ times.}}}
    \label{fig:Figure_S_Random_Circuits}
\end{figure}

\begin{table}[ht]
\centering
\resizebox{0.95\textwidth}{!}{
    \begin{tabular}{|l||c|c||c|c||c|c||c|c|}
    \hline
    \multirow{2}{*}{Qubit(s)} & \multicolumn{2}{|c||}{Q4} & \multicolumn{2}{|c||}{Q5} & \multicolumn{2}{|c||}{Q6} & \multicolumn{2}{|c|}{Q7} \\
    \cline{2-9}
	& $e_F$ & $e_U$ & $e_F$ & $e_U$ & $e_F$ & $e_U$ & $e_F$ & $e_U$ \\
    \hline
    
    \hline
    Q4 & 1.2 (0.57) & 0.16 (0.70) &  &  &  &  &  & \\
    Q5 &  &  & 1.1 (0.46) & 0.14 (0.54) &  &  &  & \\
    Q6 &  &  &  &  & 1.4 (0.35) & 0.13 (0.48) &  & \\
    Q7 &  &  &  &  &  &  & 1.9 (1.1) & 0.59 (1.1) \\
    \hline
    
    Q4 \& Q5 & 4.8 (6.9) & 3.7 (6.9) & 6.0 (9.0) & 4.9 (9.1) &  &  &  & \\
    Q4 \& Q6 & 1.1 (4.8) & 0.19 (0.71) &  &  & 2.0 (2.2) & 0.97 (2.4) &  & \\
    Q4 \& Q7 & 0.96 (0.46) & 0.23 (0.54) &  &  &  &  & 1.7 (1.0) & 0.85 (1.1) \\
    Q5 \& Q6 &  &  & 2.2 (1.6) & 1.2 (1.6) & 1.5 (0.87) & 0.40 (0.95) &  & \\
    Q5 \& Q7 &  &  & 1.5 (1.0) & 0.57 (1.1) &  &  & 2.4 (2.3) & 1.3 (2.5) \\
    Q6 \& Q7 &  &  &  &  & 4.8 (5.7) & 3.7 (5.7) & 3.9 (3.7) & 2.6 (3.7) \\
    \hline
    
    Q4, Q5, \& Q6 & 3.9 (5.1) & 2.8 (5.1) & 8.3 (13.0) & 6.7 (13.0) & 3.2 (3.1) & 1.9 (3.4) &  & \\
    Q5, Q6, \& Q7 &  &  & 3.5 (3.9) & 2.5 (3.9) & 5.7 (7.8) & 4.1 (7.9) & 4.4 (4.9) & 3.3 (5.0) \\
    \hline
    
    \end{tabular}}
\caption{
    \textbf{RB ($e_F$) and unitary RB ($e_U$) process infidelities measured before the random single-qubit circuits of variable depth experiments.} These values are used to quantify the fraction of the total error rate due to coherent errors (see Fig.~\ref{fig:Figure_S_TVD_vs_K}). All process infidelities are $\times 10^{-3}$ and all standard deviations are $\times 10^{-4}$. }
\label{tab:tvd_vs_k_process_infidelities}
\end{table}

\noindent
Random bare circuits were generated by randomly sampling $K$ interleaved cycles of easy $C_{easy}$ and hard $G_{hard}$ gates from a universal gate set (see Fig.~\ref{fig:Figure_S_Random_Circuits}(a)). For the (isolated and simultaneous) single-qubit circuits, $C_{easy} = \{ \mathbf{C_1} \}$ and $G_{hard} = \{ X45, \; Y45, \; T=Z45 \}$, where $\mathbf{C_1}$ is the single-qubit Clifford set. For multi-qubit circuits involving entangling operations, $C_{easy} = \{ \mathbf{C_1}, \; X45, \; Y45, \; T \}$ and $G_{hard} = \{ CX=CNOT, \; CY, \; CZ \}$. For easy gate cycles, single-qubit gates are randomly sampled from $C_{easy}$ independently for each qubit. For hard gate cycles in single-qubit circuits, single-qubit gates are randomly sampled from $G_{hard}$ independently for each qubit. For hard gate cycles in multi-qubit circuits, a two-qubit gate is sampled from $G_{hard}$ for a single pair of nearest-neighbor qubits, and identity gates are applied to the remaining spectator qubits. The native two-qubit gate in our system is a $CNOT$. Therefore, in order to convert the $CY$ and $CZ$ gates into our native $CNOT$ gate, the appropriate single-qubit gates were inserted around the $CX$ and then recompiled with the surrounding single-qubit gates in order to maintain the same circuit depth, before generating the experimental pulse sequences. Because the $\{ H, \; S, \; T, \; CNOT \}$ set of gates can be used for universal quantum computation, and $H$, $S$, and $CNOT$ are all Clifford gates, this is typically referred to as the universal `Clifford + $T$' set.

Once a random bare circuit has been generated, the following are the computational steps required to implement RC with $n$ total shots (Fig.~\ref{fig:Figure_S_Random_Circuits}(b)):
\begin{enumerate}
    \item Use the protocol to generate $N$ logically-equivalent randomizations of the bare circuit.
    \item For each randomization, convert the abstract gates into experimental pulse sequences and measure $n/N$ times.
    \item Compute the union of all $N$ results, such that the total number of shots over $N$ randomizations is $n$.
\end{enumerate}
Generating many randomizations requires very low classical overhead and can be efficiently done before run time. Converting $N$ randomizations into experimental pulse sequences and uploading them to the quantum computer scales linearly in $N$, but the actual measurement time is unchanged since the total number of shots $n$ remains the same as the bare circuit. In step 3, the results from all $N$ randomizations are combined into a single distribution before computing the TVD with respect to the ideal results.

\begin{figure}
\centerline{\includegraphics[width=0.65\textwidth]{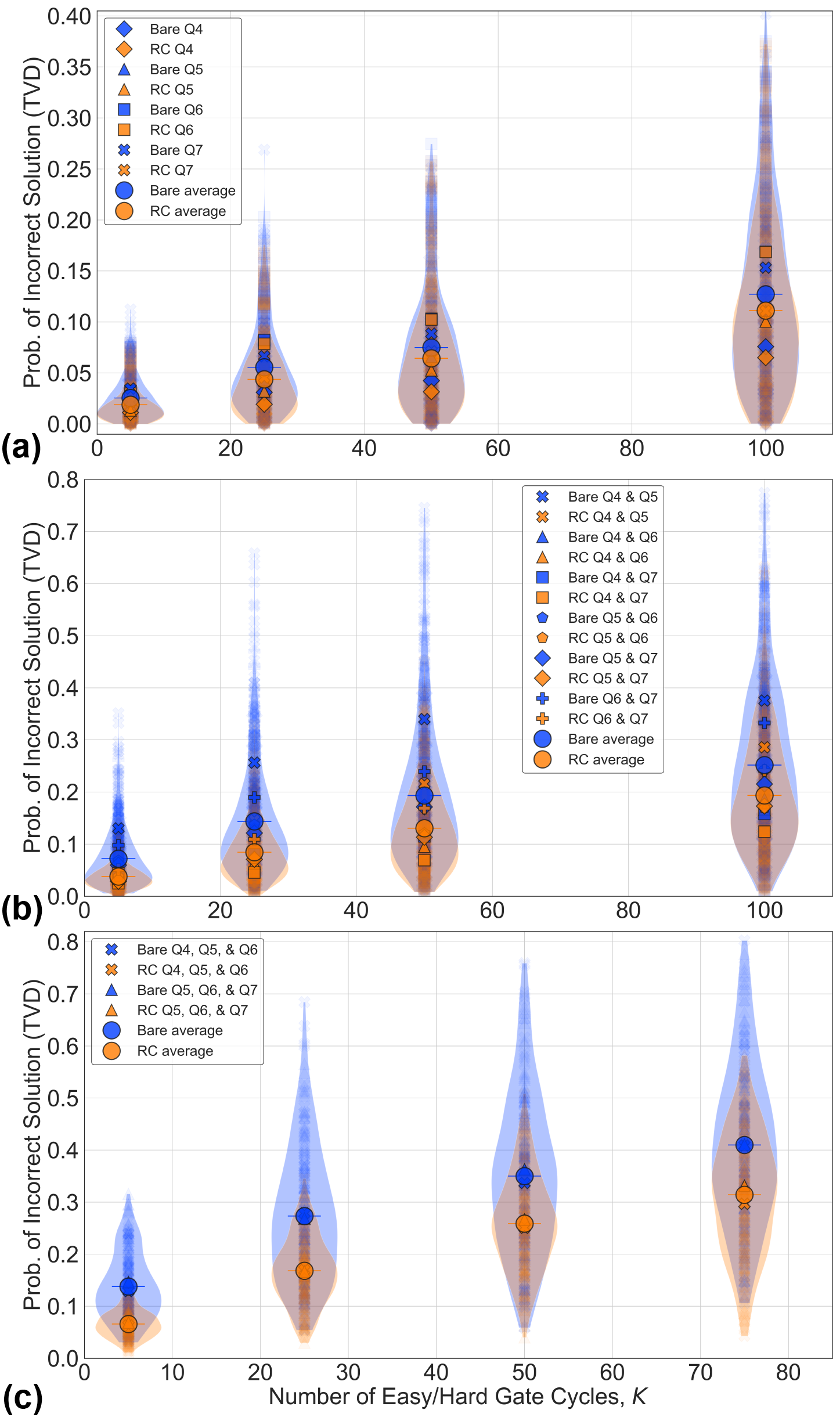}}
\caption{
    \textbf{Single-qubit random circuits of variable depth.} RB ($e_F$) and unitary RB ($e_U$) process infidelities were measured before each set of experiments to quantify the fraction of the total error rate due to coherent errors (see Table \ref{tab:tvd_vs_k_process_infidelities}).
    \textbf{(a)} \normalfont{Isolated single-qubit circuits.} 
    \textbf{(b)} \normalfont{Simultaneous single-qubit circuits on two qubits.}
    \textbf{(c)} \normalfont{Simultaneous single-qubit circuits on three qubits.} The $K=5$ data from each of the above plots was used in Fig.~\ref{fig:Figure4}(d) of the main text.}
    \label{fig:Figure_S_TVD_vs_K}
\end{figure}

In Fig.~\ref{fig:Figure_S_TVD_vs_K}, we plot random circuits of variable depth for single-qubit circuits performed in isolation or in parallel. As more qubits are operated in parallel, the error rate due to coherent errors increases due to crosstalk (see Table \ref{tab:tvd_vs_k_process_infidelities}); however, RC provides an average reduction in the TVD at all circuit depths tested. The $K=5$ data for each qubit subset in these plots was used to analyze the data in Fig.~\ref{fig:Figure4}(d) of the main text, in which the average TVD improvement factor under RC was plotted against the average fraction of the total error rate due to coherent errors. In general, for a fixed total error budget, RC performance increases as the fraction of the total error rate due to coherent errors increases.

The simulation results in Fig.~\ref{fig:Figure_S_QFT_improvement_sim_scaled_1Q} and \ref{fig:Figure_S_QFT_improvement_sim_scaled_both} show that for a fixed fraction of the total error due to coherent errors, RC performance will improve as the total error rate decreases. Therefore, while coherent errors may account for a smaller fraction of the total error budget on nearly-coherence-limited systems, and thus worse RC performance relative to systems with a larger fraction of coherent errors, RC performance may in fact still remain significant due to the relative lower overall error rates. Most importantly, our experimental and simulation results suggest that for any system in which coherent errors persist, randomized compiling will provide some improvement in algorithm performance on average.

\end{document}